# HetRoute: Heterogeneous and Cost-aware Collaborative Routing Framework for Distributed Edge MoE Inference

Xin Yuan, Ning Li, Wenchao Xu, Song Guo, *Fellow, IEEE*, Haijun Zhang, *Fellow, IEEE*

**Abstract—Mixture-of-Experts (MoE) models have become a dominant architecture for large-scale AI services, yet deploying them over geo-distributed heterogeneous edge servers remains challenging. When the Top-k activated experts of a token are spread across multiple servers, the optimal routing depends jointly on cross-server link bandwidth, heterogeneous GPU computing capability, GPU-CPU expert loading delay, instantaneous queueing backlog, and replica-level quantization quality loss. Existing distributed inference and MoE serving methods address these factors separately and do not provide a unified framework for online multi-server collaborative routing. In this paper, we propose HetRoute, a heterogeneous-cost-aware collaborative routing framework for distributed edge MoE inference. HetRoute introduces a unified per-assignment cost model that explicitly captures four cost components: cross-server transmission, GPU-CPU offloading, GPU computation with queueing, and quantization-induced quality penalty. Guided by this model, the offline stage determines expert server placement, GPU-CPU residency, and replica precision through a routing-cost-coupled deployment algorithm, while the online stage routes the Top-k activated expert set as a whole by minimizing the bottleneck layer cost via exact enumeration or beam search. Theoretical analysis establishes fallback feasibility, a bound on the number of participating servers, per-layer optimality for small candidate domains, and online computational complexity. Trace-driven evaluation on three MoE models over a heterogeneous 10-server edge testbed shows that HetRoute reduces average inference latency by up to 59.0% and P99 latency by up to 58.0%, cuts cross-server traffic by up to 72.1%, and achieves 2.13x throughput improvement compared with representative baselines, while keeping quality degradation within the configured budget.**

## I. Introduction

Large language models (LLMs) have recently shown remarkable capability in natural language understanding, code generation, multimodal perception, autonomous driving, embodied intelligence, etc. [1]-[5]. However, the rapid growth of model scale also brings extremely high computation, memory, and communication requirements. Modern LLMs usually contain billions or even hundreds of billions of parameters, and their inference requires large GPU memory, high computing capability, and frequent access to model weights and intermediate activations. Therefore, it is difficult to directly deploy such models on resource-constrained devices, such as mobile phones, vehicles, unmanned aerial vehicles (UAVs), and lightweight edge terminals. To address this issue, edge intelligence (EI) has been proposed as a promising paradigm, where AI models are compressed, partitioned, offloaded, or collaboratively deployed across devices, edge servers, and clouds to reduce the resource burden of devices and improve inference efficiency [6]-[12].

In EI, several techniques have been investigated to make LLMs more deployable under limited resources, including model compression, quantization, model splitting, edge-cloud offloading, and distributed deployment [8]-[12]. These techniques reduce the memory footprint of models or distribute different parts of a model across multiple nodes. Recently, the Mixture-of-Experts (MoE) architecture has provided a new opportunity for EI. Different from dense Transformer models that activate all parameters for every token, MoE models activate only a small subset of experts in each layer through Top-k sparse routing [13]-[15]. This sparse activation property naturally matches the resource-limited nature of EI: only the activated experts need to participate in inference, while the remaining experts can stay inactive. Based on this property, recent studies have explored MoE deployment on edge or resource-constrained devices from different perspectives, such as GPU-CPU collaborative expert placement, expert caching and offloading, model quantization, distributed expert deployment, etc. [16]-[23].

Although these studies improve the feasibility of deploying MoE models outside centralized cloud data centers, they still cannot fully address the problem of distributed collaborative MoE inference over heterogeneous edge servers. Because when an MoE model is deployed over multiple edge servers in a distributed collaborative manner, several new challenges arise. First, the Top-k activated experts of one token may be distributed over multiple edge servers. Due to redundant deployment [38], the candidate collaboration domain may even contain more servers than the number of activated experts. Thus, token routing is no longer a single-server selection problem, but a multi-server collaborative routing problem. Second, experts are usually deployed cooperatively between

This paragraph of the first footnote will contain the date on which you submitted your paper for review, which is populated by IEEE. This work was supported in part by the grant from NSFC Grant no. 62571156, 62101159, 52475009, NSF of Shandong Grant no. ZR2021MF055, ZR2025QC666, the Research Grants Council of Hong Kong under the Areas of Excellence scheme grant AoE/E-601/22-R, and also the Opening Project of the Key Laboratory of Advanced Manufacturing and Intelligent Technology (Ministry of Education) at Harbin University of Science and Technology (KFKT202306).
*(Corresponding author: Ning Li).*

Ning Li and Haijun Zhang are with the School of Artificial Intelligence, University of Science and Technology Beijing, China (e-mail: ningli_polyuhk@outlook.com, zhanghaijun@ustb.edu.cn).

Xin Yuan is with the School of Ocean Engineering, Harbin Institute of Technology, Heilongjiang, China (e-mail: xin.yuan@hit.edu.cn).

Wenchao Xu is with the Division of Integrative Systems and Design, Hong Kong University of Science and Technology, Hong Kong (e-mail: wenchaoxu@ust.hk)

Song Guo is with the Department of Computer Science and Engineering, Hong Kong University of Science and Technology, Hong Kong (e-mail: songguo@cse.ust.hk)

Mentions of supplemental materials and animal/human rights statements can be included here.

Color versions of one or more of the figures in this article are available online at http://ieeexplore.ieee.org

GPU and CPU inside each server: hot experts are kept in GPU memory to reduce execution latency, while cold experts are stored in CPU memory to save GPU memory. In this case, a locally available cold expert may still incur significant GPU-CPU loading delay, and may be slower than a remote hot expert that is already resident in GPU memory. Third, the inter-server links in edge networks are ordinary and heterogeneous Internet links rather than stable high-speed data center connection technology, such as RDMA, InfiniBand, etc. The bandwidth and queueing latency can be significantly different across server pairs. Fourth, to fit more expert replicas into limited edge memory, different replicas may be stored with different quantization precisions, which introduces different quality losses. These challenges indicate that local-first routing, placement-only optimization, and per-expert greedy selection are insufficient for distributed edge MoE inference.

Fig. 1 illustrates a simple example. A token may activate two target experts in one MoE layer. One edge server stores both experts, but the experts are cold experts residing in CPU memory and must be loaded into GPU before execution. Two other edge servers store one activated expert each, and both experts are hot experts already residing in GPU memory. In this case, executing the token locally avoids cross-server communication, but suffers from GPU-CPU loading delay and possibly with weak GPU computation. Executing the two experts on two remote servers incurs fan-out and fan-in transmissions, but may achieve lower overall latency because the remote experts are GPU-resident and the remote servers have higher computing capability. Therefore, the optimal route depends jointly on link bandwidth, GPU capability, GPU-CPU loading cost, server backlog, and quality loss. A routing strategy that only considers expert location or selects the individually best server for each expert may choose a locally attractive but globally expensive collaboration path.

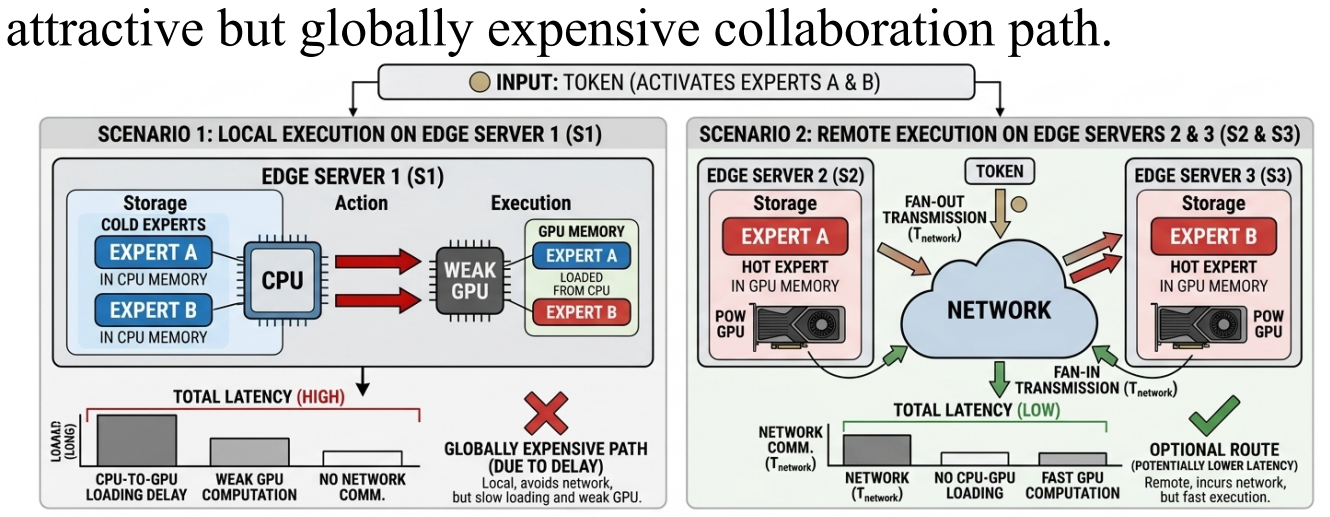

Fig.1 An Example

Based on the above analysis, a heterogeneous and cost aware collaborative routing framework for distributed edge MoE inference is proposed in this paper, abbreviated as HetRoute. Specifically, HetRoute considers an edge inference system with multiple heterogeneous edge servers serving MoE inference requests. In the offline stage, HetRoute determines where expert replicas are deployed, whether each expert is placed in GPU or CPU memory, and which quantization precision is used for each replica. Instead of relying only on raw expert activation frequency, the offline deployment is coupled with the online routing objective: the GPU residency benefit and redundant-replication benefit are evaluated according to the expected online unified cost under calibration traffic. In the online stage, HetRoute routes each token layer by layer. For each Top-k activated expert set, HetRoute constructs a candidate collaboration domain from exact and feasible substitute replicas, prunes candidates according to quality and stability guards, and then selects a complete collaboration set by minimizing the layer-level bottleneck cost. In this way, HetRoute routes the Top-k experts as a set rather than as independent per-expert decisions.

The main contributions of this paper can be summarized as follows.

1) To the best of our knowledge, this is the first work that formulates distributed edge MoE inference as a heterogeneous cost-aware Top-k collaborative routing problem. Different from existing EI and MoE inference algorithms, HetRoute jointly models cross-server transmission, heterogeneous GPU computation, GPU-CPU expert loading, queueing backlog, and replica-level quantization quality loss. This unified model captures the important case that a remote GPU-resident hot expert may be faster than a local CPU-resident cold expert.

2) We propose a set-level collaborative routing algorithm for Top-k activated MoE experts. Instead of greedily selecting the best server for each expert independently, HetRoute evaluates the bottleneck delay of the complete collaboration set, including fan-out transmission, parallel expert execution, fan-in aggregation, GPU-CPU loading delay, and quality cost. This design explicitly captures the coupling among multiple activated experts and avoids globally inefficient multi-branch routing decisions.

3) We design an offline-online co-optimization strategy for expert deployment, GPU-CPU residency, and replica precision. The offline stage uses calibration traffic and the online unified cost to estimate replica-selection probabilities, GPU-residency benefits, and redundancy benefits. Thus, the generated expert deployment is not a static placement result, but a routing-cost-coupled candidate structure that supports efficient online collaborative routing.

4) We analyze the feasibility, structural property, optimality scope, and complexity of HetRoute, and evaluate it through trace-driven experiment on three MoE models. The results show that HetRoute significantly reduces average latency, tail latency, cross-server traffic, and CPU-offload ratio, while improving throughput and keeping the quality degradation within the configured budget.

The remainder of this paper is organized as follows. Section II reviews the related works. Section III presents the network model, the distributed MoE deployment model, the unified heterogeneous cost model, and the problem formulation. Section IV introduces the offline deployment algorithm. Section V presents the online collaborative routing algorithm and its theoretical properties. Section VI evaluates the performance of HetRoute. Finally, Section VII concludes this paper.

## II. Related Works

In this section, the details of the most related works are introduced. The related works can be divided into three categories: collaborative inference and distributed deployment of large models at the edge, MoE serving and expert offloading, and model partitioning and memory-aware offloading.

### *A. Collaborative inference and distributed deployment of large models at the edge*

Research on collaborative inference and distributed deployment primarily aims to deploy large models over multiple resource-constrained devices, edge servers, or cloud servers, thereby reducing the resource burden of individual devices and improving inference latency [6]-[12]. EdgeShard [7] proposes a collaborative edge LLM inference framework that partitions dense LLMs across heterogeneous edge devices and cloud servers. It jointly considers device selection and model partitioning to improve inference latency and throughput. Petals [22] enables collaborative inference and fine-tuning of large models over decentralized nodes, where different participants hold different model blocks and users execute inference through networked peers. These studies demonstrate that large models can be served collaboratively without requiring one device to hold the entire model.

However, these collaborative inference systems are mainly designed for dense Transformer models or block-wise model partitioning. The inference path of dense models is usually serial at the layer or block level, while MoE models activate only a sparse subset of experts at each layer. As a result, the main decision in HetRoute is not where to cut a dense model, but how to route a token's Top-k activated experts to a collaboration set of servers. Moreover, existing collaborative inference systems generally do not jointly consider expert-level redundancy, GPU-CPU expert residency, replica-level quantization precision, and substitute execution. Therefore, directly applying dense-model collaborative inference to distributed edge MoE serving cannot fully exploit the sparse expert activation property.

Recently, several works have started to investigate LLM inference at the edge. For example, EdgeMoE studies sparse large language model deployment on edge devices and explores the potential of MoE models under edge constraints [24]. D2MoE investigates distributed dynamic inference for MoE models under edge resource limitations [25]. These studies indicate that MoE is a promising architecture for edge intelligence because sparse activation reduces the number of parameters used per token. Nevertheless, they do not fully solve the heterogeneous multi-server routing problem considered in this paper. HetRoute differs from them by modeling the candidate collaboration domain of each Top-k expert set and by optimizing the complete set-level bottleneck routing cost online.

### B. MoE serving, expert placement, and expert offloading

MoE serving systems mainly focus on the sparsity and expert activation patterns of MoE models. Since only a small number of experts are activated for each token, the system can reduce memory and communication overhead by predicting, caching, prefetching, or offloading experts [16]-[21]. MoE-Infinity [16] is an activation-aware MoE serving system that traces expert activation and performs expert caching and prefetching between host memory and GPU memory. SiDA-MoE [21] further exploits the sparsity and data-aware characteristics of MoE inference to improve scalable serving of large MoE models. These works show that expert activation patterns can be used to reduce GPU memory pressure and improve inference efficiency.

In addition to memory offloading, expert placement is also critical for distributed MoE inference. Prism [23] is a closely related work that accelerates distributed edge MoE inference through latency-optimized expert placement. It places experts across edge servers according to activation and latency information, so that cross-server communication can be reduced. Other MoE systems such as Tutel and MegaBlocks improve MoE training and serving efficiency through optimized expert parallelism, token dispatching, and sparse computation kernels [17], [18]. These systems greatly improve MoE execution efficiency in data-center or GPU-cluster environments.

However, the above MoE serving and placement methods still have several limitations when they are used in heterogeneous edge networks. First, many expert offloading systems mainly focus on a single server or a tightly coupled GPU-host memory hierarchy, and do not model ordinary Internet links among edge servers. Second, expert placement alone cannot adapt to the instantaneous online state, such as time-varying link bandwidth, GPU backlog, and whether the selected expert is in GPU or CPU memory. Third, existing methods usually optimize each expert or placement decision separately, while the Top-k activated experts in one MoE layer interact through fan-out transmission, parallel execution, and fan-in aggregation. HetRoute addresses these limitations by coupling offline expert deployment with online set-level collaborative routing. It uses the offline stage to construct a useful collaboration domain, and uses the online stage to select a complete collaboration set under the unified heterogeneous cost model.

### C. Model partitioning and memory-aware offloading

Model partitioning and offloading have been studied for many years in edge intelligence and mobile-cloud computing [26]-[33]. Neurosurgeon [26] profiles DNN layers and partitions neural network inference between mobile devices and cloud servers. DDNN [27] distributes deep neural networks across cloud, edge, and end devices with early exits. JointDNN [28] supports collaborative computation between mobile devices and cloud for DNN inference and training. DADS [29] dynamically partitions DNNs according to network conditions, and other works further consider multi-user resource allocation, early exit, and intermediate feature compression [30]-[33]. These studies provide important foundations for edge intelligence and show that model partitioning can reduce latency and resource consumption.

For large language models, memory-aware offloading has become an important research direction. FlexGen [8] improves high-throughput LLM inference on a single GPU by using GPU, CPU, and disk memory together and optimizing tensor placement and access patterns. Recent quantization methods such as QLoRA, GPTQ, SmoothQuant, and AWQ reduce the memory usage of LLMs by using low-bit representations while preserving model quality [34]-[37]. These methods are relevant to HetRoute because GPU-CPU memory movement and precision selection are also important factors in distributed edge MoE inference.

However, model partitioning and memory-aware offloading cannot directly solve the problem addressed by HetRoute. Traditional partitioning methods usually assume a dense model execution path and focus on finding a split point or a pipeline schedule. FlexGen-like systems mainly optimize the memory hierarchy of a single serving node, rather than routing Top-k sparse experts over multiple heterogeneous edge servers. Quantization methods reduce memory usage but usually do not decide where each quantized expert should be placed or how a token should be routed under a per-token quality budget. In contrast, HetRoute integrates these factors into a unified framework: expert replicas may be placed on different servers with different precisions, each expert may reside in GPU or CPU memory, and the online router chooses a collaboration set according to transmission, computation, offloading, queueing, and quality costs.

*D. Summary of differences*

In summary, existing collaborative inference works mainly optimize dense model partitioning, existing MoE serving works mainly optimize expert caching, prefetching, or placement, and existing offloading works mainly optimize memory movement within a single node or a simple edge-cloud path. These works do not fully address the joint problem of heterogeneous edge MoE inference, where Top-k activated experts may have multiple exact or substitute replicas across heterogeneous servers, and where the best route depends on cross-server bandwidth, GPU capability, GPU-CPU residency, queueing backlog, and quantization-induced quality loss. HetRoute fills this gap by proposing a unified heterogeneous cost model, an offline routing-cost-coupled deployment algorithm, and an online set-level collaborative routing algorithm for distributed edge MoE inference.

## III. Network Model and Problem Statement

In this section, we introduce the details of the network model, the distributed MoE deployment, the unified heterogeneous cost model, and the quality degradation model, which are integral to this paper. Following this, the specific problem addressed in our study, guided by the proposed models, is thoroughly outlined.

*A. Network Model and Distributed MoE Deployment*

This paper examines a distributed edge intelligence system that serves MoE inference, comprising $N$ heterogeneous edge servers and $U$ end users, labeled as $\mathcal{N}=\{s_1,s_2,\dots,s_N\}$ and $\mathcal{U}=\{1,2,\dots,U\}$, respectively. The edge servers are geographically distributed and interconnected through ordinary Internet links. For any server pair $(s_m,s_n)$, we use $B_{\text{link}}^{m,n}$ to denote the available link bandwidth from $s_m$ to $s_n$, which differs across server pairs. Each server $s_n\in\mathcal{N}$ is heterogeneous in both computation and memory: $F_n$ (FLOPs/s) denotes its GPU computing capability, $\beta_n$ denotes its GPU-CPU data-transfer (e.g., PCIe) bandwidth, and $G_n$ and $C_n$ denote its GPU memory capacity and CPU memory capacity, respectively. For each user $u\in\mathcal{U}$, its associated home server $a_u\in\mathcal{N}$ receives the initial inference request and emits the final output. Unlike conventional designs in which the home server also aggregates the result of every layer, in this paper the per-layer aggregation server varies dynamically along the token trajectory, as detailed in Section III-D.

The served model is an MoE model consisting of $L$ MoE layers, labeled as $\mathcal{L}=\{1,2,\dots,L\}$. Each layer $l\in\mathcal{L}$ contains an expert set $\mathcal{E}_l$, and every expert $E\in\mathcal{E}_l$ is associated with a computation load $w_E$ (FLOPs) and a similarity-based feasible substitute set $\mathcal{V}_{l,r_l(i)}$ consistent with the substitute execution model in [38]. Unlike conventional model partitioning works in which a layer is statically placed on a single node, the placement of an expert in this paper has two coupled dimensions: on which server the expert (or its redundant replica) resides, and in which memory tier, i.e., GPU or CPU, the replica is held. Accordingly, we introduce two offline deployment variables:

$$x_{E,n}=\begin{cases}1, \text{Expert } E \text{ is deployed on server } s_n\\ 0, \quad \text{otherwise}\end{cases} \tag{1}$$

$$y_{E,n}=\begin{cases}1, \text{the deployed replica of } E \text{ on } s_n \text{ resides in GPU}\\ 0, \quad \text{otherwise}\end{cases} \tag{2}$$

Every deployed replica resides in exactly one memory tier: GPU when $y_{E,n}=1$, and CPU when $x_{E,n}=1$ and $y_{E,n}=0$. The case $x_{E,n}=0$ means that no replica of $E$ exists on $s_n$ and does not imply CPU placement. Accordingly, the constraint $y_{E,n}\le x_{E,n}$ prevents the degenerate case of a GPU-resident replica without deployment.

Following the redundant deployment principle of [38], an expert may be replicated on more than one server, i.e., $\sum_n x_{E,n}\ge 1$, so that a token does not always have to be routed to a single fixed server. For each expert $E$, its candidate-server set is defined as:

$$\mathcal{S}_E=\{s_n\in\mathcal{N}\mid x_{E,n}=1\} \tag{3}$$

Furthermore, each replica carries a server specific quantization level $c_{E,n}\in Q=\{\text{INT4},\text{INT8},\text{FP16},\dots\}$, and the memory footprint of replica $(E,N)$ is $v_{E,n}=v(c_{E,n})$. This captures the important design dimension that the same expert may be stored at high precision on one server and low precision on another, trading memory consumption against inference quality. The quality loss induced by quantization level $c_{E,n}$ is denoted $q_{E,n}=q(c_{E,n})$ and is defined in Section III-C.

*B. Top-k Routing, Collaboration Domain, and Substitute Execution*

At layer $l$, the gating network of token $\tau_i$ activates the Top-k target expert set $\mathcal{K}_{i,l}$. For each target expert $E\in\mathcal{K}_{i,l}$, HetRoute follows an exact-first substitute policy: the router first attempts to serve $E$ using one of its own deployed replicas; a feasible substitute $\hat{E}\neq E$ is considered only when the exact candidates of $E$ are unavailable or become inadmissible under the current online admission guards, for example, because all exact replicas would violate the per-token quality budget, the normal per-server stability limit, or the current reachability condition.

To model exact and substitute execution in a unified manner, we introduce the routing variable $z_{i,l,E,\hat{E},n}\in\{0,1\}$, which equals 1 if and only if target expert $E$ activated by token $\tau_i$ at layer $l$ is executed by replica $\hat{E}$ on server $s_n$. The case $\hat{E}=E$ corresponds to exact execution and incurs zero substitution loss ($Q_{\text{sub}}(E,E)=0$) and $Q_{\text{sub}}(E,\hat{E})$ is the substitution quality loss defined in [38]; the case $\hat{E}\neq E$ corresponds to substitute execution and incurs the profiled substitution quality loss $Q_{\text{sub}}(E,\hat{E})>0$. The candidate set of target expert $E$ is defined at two levels. The exact candidate set contains all servers that host a replica of $E$: $\mathcal{C}_E^{ex}=\{(E,n)\mid x_{E,n}=1\}$. The substitute candidate set contains all feasible substitute replicas from the similarity group $\mathcal{V}_E$: $\mathcal{C}_E^{sub}=\{(\hat{E},n)\big|\hat{E}\in\mathcal{V}_E,\hat{E}\neq E,x_{\hat{E},n}=1\}$. During routing, HetRoute first evaluates $\mathcal{C}_E^{ex}$. The substitute set $\mathcal{C}_E^{sub}$ is activated only if $\mathcal{C}_E^{ex}$ contains no admissible assignment after the online admission guards are applied. The active candidate set used by the online router is therefore:

$$\mathcal{C}_E=\begin{cases}\mathcal{C}_E^{ex}, \text{if } \mathcal{C}_E^{ex} \text{ contains at least one admissible candidate}\\ \mathcal{C}_E^{sub}, \text{otherwise}\end{cases} \tag{5}$$

This exact-first policy removes the ambiguity between exact replica routing and similarity-based substitute execution: $\mathcal{K}_{i,l}$ always denotes the target experts selected by the gating network, while $\hat{E}$ denotes the actual executed replica, which may be either the target itself or a feasible substitute.

It is important to distinguish the candidate collaboration domain from the actual participating set in order to clarify the relationship between redundancy and routing. The candidate collaboration domain of token $\tau_i$ at layer $l$ is the union of all candidate servers over the activated experts:

$$\mathcal{C}_{i,l} = \cup_{E \in \mathcal{K}_{i,l}} \{n \mid (\hat{E}, n) \in \mathcal{C}_E\} \quad (6)$$

where $|\mathcal{C}_{i,l}|$ may exceed $k$. Whereas the participating set of servers actually selected to execute layer $l$ is:

$$\mathcal{P}_{i,l} = \{n \mid \exists\, E, \hat{E}: z_{i,l,E,\hat{E},n} = 1\}, |\,\mathcal{P}_{i,l}\,| \leq k \quad (7)$$

Consequently, the collaboration domain can span many servers due to redundant deployment, while at most $k$ servers ever participate in a given layer. This distinction directly resolves the apparent inconsistency between redundancy and the Top-$k$ constraint: the domain size may exceed $k$ because each expert has multiple replica candidates, but each target expert is executed exactly once, so the number of participating servers is bounded by $k$.

*C. Quality Degradation Model*

The quantization loss $q_{E,n}$ is obtained by offline profiling rather than by assumption. Specifically, for each quantization level $c \in Q$, we measure the mean increase of validation loss (KL divergence from full-precision output) on the calibration set $\mathcal{T}_{\text{cal}}$ when the replica of $E$ is stored at precision $c$, and record $q_{E,n} = q(c_{E,n})$ accordingly. This ensures that the quality model reflects the actual precision configuration of each individual replica.

The accumulated quality degradation of token $\tau_i$ is then expressed as the additive surrogate:

$$\Lambda_i = \sum_{l=1}^{L} \sum_{E \in \mathcal{K}_{i,l}} \sum_{(\hat{E},n)} z_{i,l,E,\hat{E},n} \left(q_{\hat{E},n} + Q_{\text{sub}}(E, \hat{E})\right) \quad (8)$$

where $q_{\hat{E},n}$ is the profiled quantization loss of the executed replica. The additive form of (8) is stated as a first-order approximation of the end-to-end quality loss. Its correlation with the measured task metric (e.g., perplexity or accuracy) is validated empirically in Section VI.

For consistency throughout the system, we adopt two normalizations. First, the quantization loss of a full-precision replica is defined to be zero: $q_{E,n} \triangleq 0$, whenever $c_{E,n} = c_{\max}$. Second, exact execution incurs no substitution penalty: $Q_{\text{sub}}(E, E) \triangleq 0$. These two normalizations together imply that a full-precision exact replica always contributes zero quality degradation to the accumulated budget, which is the foundational premise of the emergency fallback mechanism formalized in Section III-E (c.4) and proved in Property 1.

*D. Unified Heterogeneous Cost Model*

Let $\theta_{i,l}$ denote the residing server of token $\tau_i$ before layer $l$, with $\theta_{i,1} = a_i$. Executing target expert $E$ via replica $\hat{E}$ on server $s_n$ incurs four heterogeneous cost components. First, if $n \neq \theta_{i,l}$, the hidden-state representation of size $d_i^{\text{in}}$ must be transmitted over the link $\theta_{i,l} \to s_n$. Second, if the replica is CPU-resident ($y_{\hat{E},n} = 0$), it must be loaded into GPU memory before computation. Third, the GPU computation itself takes time proportional to $w_{\hat{E}}$ and is subject to queueing due to concurrent requests. Fourth, the quantization and substitution choices introduce a quality penalty. These four components are gathered into the unified per-assignment cost:

$$C_{i,l}(E, \hat{E}, n) = \underbrace{\frac{d_i^{\text{in}}}{B_{\text{link}}^{\theta_{i,l},n}} \mathbb{1}[n \neq \theta_{i,l}]}_{\text{trasnmission}} + \underbrace{\frac{(1-y_{\hat{E},n})\, v_{\hat{E},n}}{\beta_n}}_{\text{offload}} + \underbrace{\left(\frac{w_{\hat{E}} + Q_n(t)}{F_n}\right)}_{\text{compute+queue}} + \lambda_q \left(q_{\hat{E},n} + Q_{\text{sub}}(E, \hat{E})\right) \quad (9)$$

where $Q_n(t)$ is the GPU backlog on $s_n$ (in FLOPs) already admitted in scheduling window $t$, so $Q_n(t)/F_n$ captures the instantaneous queueing delay that the token actually experiences. The transmission term vanishes when the replica is co-located with the residing server. Consequently, local execution is a zero-transmission candidate rather than a default optimum.

The above model indicates that the server selection, replica selection, and substitute execution decisions are jointly related to both inference delay and quality degradation. For example, routing an activated expert to a remote server whose replica is GPU resident reduces computation-plus-loading delay but incurs cross-server transmission delay; executing it locally on a CPU resident replica removes the transmission delay but introduces an offload penalty. This trade-off motivates a unified cost framework rather than any local-first or remote-first heuristic.

Because the $k$ activated experts run in parallel on their respective participating servers, the layer delay is determined by the slowest branch, i.e., it is a bottleneck (max) rather than a sum. Specifically, the layer-$l$ delay of token $\tau_i$ consists of a fan-out phase, a parallel compute phase, and a fan-in phase:

$$D_{i,l} = \underbrace{\max_{n \in \mathcal{P}_{i,l} \setminus \{\theta_{i,l}\}} \frac{d_i^{\text{in}}}{B_{\text{link}}^{\theta_{i,l},n}}}_{\text{fan-out}} + \underbrace{\max_{n \in \mathcal{P}_{i,l}} \left(\frac{(1-y_{\hat{E},n})\, v_{\hat{E},n}}{\beta_n} + \frac{w_{\hat{E}} + Q_n(t)}{F_n}\right)}_{\text{slowest compute branch}} + \underbrace{\max_{n \in \mathcal{P}_{i,l} \setminus \{\theta_{i,l+1}\}} \frac{d_i^{\text{out}}}{B_{\text{link}}^{n,\theta_{i,l+1}}}}_{\text{fan-in}} \quad (10)$$

Here, $d_i^{\text{out}}$ is the partial-output size per branch, and the fan-in delay is measured to the aggregation server $\theta_{i,l+1}$. Consequently, the end-to-end inference delay of token $\tau_i$ is $D_i^{\text{tot}} = \sum_{l=1}^{L} D_{i,l}$.

After all $k$ partial outputs are gathered, the token resides on the aggregation server $\theta_{i,l+1}$, which is selected to minimize the gather traffic plus a one-step lookahead of the next layer's routing cost:

$$\theta_{i,l+1} = \arg\min_{n \in \mathcal{P}_{i,l}} \left[\sum_{m \in \mathcal{P}_{i,l} \setminus \{n\}} \frac{d_i^{\text{out}}}{B_{\text{link}}^{m,n}} + \hat{T}_{i,l+1}^{\text{next}}(n)\right] \quad (11)$$

where $\hat{T}_{i,l+1}^{\text{next}}(n)$ is a one-step lookahead estimate of the cost incurred in the next layer if the token resides on $n$. We note that (11) is a heuristic selection rule; it is not claimed to minimize the total multi-layer cost. When $\hat{T}_{i,l+1}^{\text{next}} \equiv 0$, (11) degenerates to pure gather-cost minimization. The token trajectory $\{\theta_{i,l}\}_{l=1}^{L}$ is thus determined layer by layer, with $\theta_{i,1} = a_i$ as the sole fixed point.

*E. Problem Formulation*

From the above subsections, we obtain the end-to-end inference delay $D_i^{\text{tot}}$ and the accumulated quality degradation $\Lambda_i$ of token $\tau_i$. The purpose of this paper is to minimize the inference delay and the quality degradation by jointly optimizing the expert deployment strategy $\mathbf{X}=\{x_{E,n}\}$, the residency strategy $\mathbf{Y}=\{y_{E,n}\}$, and the routing strategy $\mathbf{Z}=\{z_{i,l,E,\hat{E},n}\}$. Since the inter-server links and the GPU resources are shared, the per-server workload within a scheduling slot $\Delta t$ is bounded. Let $\mathcal{R}(t)$ denote the set of tokens served in slot $t$. The ideal joint optimization problem **P0** is then expressed as:

$$\textbf{P0}: \min_{\boldsymbol{X},\boldsymbol{Y},\boldsymbol{Z}} \sum_{\tau_i\in\mathcal{R}(t)} \left(D_i^{\text{tot}} + \lambda\,\Lambda_i\right)$$

s.t.

$$\sum_E v_{E,n}\cdot y_{E,n} \le G_n,\ \ \forall n \quad \text{(c.1)}$$

$$\sum_E v_{E,n}\cdot x_{E,n}\cdot\left(1-y_{E,n}\right) \le C_n,\ \ \forall n \quad \text{(c.2)}$$

$$y_{E,n} \le x_{E,n}, \forall E,n \quad \text{(c.3)}$$

$$\exists n: x_{E,n}=1 \wedge c_{E,n}=c_{max}, \forall E \quad \text{(c.4)}$$

$$x_{E,n}, y_{E,n}, z_{i,l,E,\hat{E},n} \in \{0,1\} \quad \text{(c.5)}$$

$$z_{i,l,E,\hat{E},n} \le x_{\hat{E},n}, \hat{E}\in\{E\}\cup\mathcal{V}_{l,r_l(i)}, \forall i,l,E,\hat{E},n \quad \text{(c.6)}$$

$$\sum_{\hat{E},n} z_{i,l,E,\hat{E},n} = 1, \forall i,l,E\in\mathcal{K}_{i,l} \quad \text{(c.7)}$$

$$|\,\mathcal{P}_{i,l}\,| \le k, \forall i,l \quad \text{(c.8)}$$

$$\Lambda_i \le \Lambda_i^{\max}, \forall i \quad \text{(c.9)}$$

$$\sum_{\tau_i\in\mathcal{R}(t)}\sum_l\sum_{E,\hat{E}} z_{i,l,E,\hat{E},n}\cdot w_{\hat{E}} \le F_n\,\Delta t, \forall n \quad \text{(c.10)}$$

In Problem P0, the constraints (c.1) and (c.2) bound the GPU and CPU memory consumption on each server, respectively. Constraint (c.3) ensures that a GPU-resident replica cannot exist without a corresponding deployment, and together with (c.1) to (c.2) it enforces that every deployed replica resides in exactly one memory tier. Constraint (c.4) is the full-precision exact-replica invariant. It requires that for every expert $E$, at least one server $s_n$ hosts an exact replica of $E$ at full precision $c_{\max}$. This invariant is strictly stronger than merely requiring $\sum_n x_{E,n}\ge 1$: it simultaneously guarantees 1) exact execution, so $Q_{\text{sub}}(E,E)=0$, and 2) full precision, so $q_{E,n}=0$ by the normalization in Section III-C. The combination gives a zero-quality increment for any assignment to this replica. Therefore, (c.4) guarantees the existence of one exact, full-precision, zero-quality-loss fallback replica for each expert, the structural foundation that makes the emergency fallback permanently quality-safe. Constraint (c.5) specifies the binary nature of all decision variables. Constraint (c.6) ensures that a routing assignment can only be made to a server that actually hosts the corresponding replica. Constraint (c.7) enforces that each target expert is executed exactly once per layer, by either its own replica or a feasible substitute. Constraint (c.8) limits the number of participating servers per layer to at most $k$, consistent with (7). The quality term appears both in the objective as $\lambda\Lambda_i$ and in the hard constraint (c.9). The constraint (c.9) is the per-token QoS bound that must never be violated, while the objective further reduces degradation when budget slack exists; the two are complementary rather than redundant. Finally, (c.10) is a per-window stability bound in which $w_{\hat{E}}$ is in FLOPs and $F_n$ is in FLOPs/s, ensuring dimensional consistency; it prevents unbounded backlog growth, while the instantaneous queueing latency experienced by the token is captured separately by the $Q_n(t)/F_n$ term in (9) and (10).

We distinguish the long-term stability constraint (c.10) from the online admission guard (will be used by Algorithm 2). Constraint (c.10) is a scheduling window stability condition enforced at the placement stage (will be introduced in Algorithm 1): it prevents the system from persistently admitting more computation than the servers can process. During normal online routing, Algorithm 2 enforces a stricter per-assignment stability guard (will be stated precisely in Section V): a candidate is admitted only if the corresponding assignment does not push the current window load beyond capacity, keeping the common routing path latency sensitive. The emergency fallback follows a different priority: if all normally admissible candidates for a target expert are exhausted, HetRoute routes that target to the full-precision exact fallback replica guaranteed by (c.4). In this exceptional case, the quality guard is never relaxed, but the normal stability guard may be relaxed, the assignment is admitted into the server queue, and the resulting overload manifests as increased queueing delay captured by the $Q_n(t)/F_n$ term in the unified cost model rather than as routing infeasibility. Thus, quality feasibility is a hard guarantee, online stability is enforced on the normal path and converted into queueing delay on the fallback path, and Algorithm 2 always returns a complete routing without violating the per-token quality budget.

This problem is difficult to solve for the following reasons. First, the objectives are conflicting: routing an activated expert to a remote server whose replica is GPU-resident reduces the computation-plus-loading delay but introduces cross-server transmission delay, while executing it on the home server whose replica is CPU-resident removes the transmission delay at the cost of an extra GPU-CPU offloading delay. Second, the deployment, residency, and routing decisions are all binary, which makes the joint solution space extremely large. Third, the decisions of different layers are coupled along the token trajectory: the execution servers selected in one layer determine the residing server from which the next layer must scatter and gather the token representation, so a per-layer routing choice directly reshapes the transmission cost of all subsequent layers. Fourth, the activated expert set $\mathcal{K}_{i,l}$ and the instantaneous link states and GPU backlogs are only revealed when the token reaches that layer, so the routing decisions must be made in an online manner. Therefore, finding the optimal solution of P0 is in general intractable.

Since P0 jointly couples the offline placement of $(\mathbf{X},\mathbf{Y})$ with the online causal routing of $\mathbf{Z}$, it cannot be solved at once in real time and serves as a non-causal ideal benchmark. We therefore decompose it into two stages: Algorithm 1 fixes $(\mathbf{X},\mathbf{Y})$ offline to minimize the expected online unified cost under the calibration traffic, and Algorithm 2 solves $\mathbf{Z}$ online per token-layer by minimizing the bottleneck layer cost (10). HetRoute is thus a two-stage decomposition heuristic for P0 rather than an exact joint solver.

Since the two objectives in P0 are of different natures, a weight-based approach is introduced to combine them into a single scalar. Let $\widetilde{D}_i = D_i^{\text{tot}}/D_i^{\text{ref}}$ and $\widetilde{\Lambda}_i=\Lambda_i/\Lambda_i^{\max}$ denote the normalized inference delay and the normalized quality degradation of token $\tau_i$, respectively, where $D_i^{\text{ref}}$ is the inference delay under the pure home-server full-precision exact-execution strategy on the same deployment. The utility function is:

$$U(\mathbf{X},\mathbf{Y},\mathbf{Z}) = \omega_T\sum_{\tau_i\in\mathcal{R}(t)}\widetilde{D}_i + \omega_\Lambda\sum_{\tau_i\in\mathcal{R}(t)}\widetilde{\Lambda}_i \quad (12)$$

where $\omega_T$ and $\omega_\Lambda$ are the weights of inference delay and quality degradation, respectively, and $\omega_T+\omega_\Lambda=1$. Thus, P0 is transformed into the single-objective problem:

$$\mathbf{P1}: \min_{\mathbf{X},\mathbf{Y},\mathbf{Z}} U(\mathbf{X},\mathbf{Y},\mathbf{Z})$$

According to (12), a larger $\omega_T$ favors low latency through more aggressive routing to high capability, GPU resident, high bandwidth servers, while a larger $\omega_\Lambda$ favors quality preservation through more exact and full-precision execution. Therefore, the weight-based approach can flexibly achieve the desired tradeoff between inference delay and inference quality.

## IV. Offline Exit-Head Training and Communication Aware Deployment

The offline stage fixes the deployment $\mathbf{X}$, the GPU residency $\mathbf{Y}$, and the per-replica quantization levels $\{c_{E,n}\}$ once per scheduling epoch. Its purpose is not merely to produce a feasible placement but to shape the candidate collaboration domain that the online router will later exploit. To make the coupling explicit and to answer the concern that offline and online stages might be two disconnected modules, every offline benefit term is defined as an expectation of the same online unified cost $C_{i,l}(\cdot)$ in (11), evaluated over the calibration traffic $\mathcal{T}_{\text{cal}}$ under the online routing policy of Algorithm 2.

Concretely, the offline stage anticipates the online cost in three coupled ways. First, an expert is given GPU residency in proportion to the offload latency that it is expected to save when it is actually selected online, so residency follows the online selection distribution rather than raw activation frequency. Second, a redundant replica is created on a server only when its marginal placement reduces the expected online unified cost of serving the corresponding target expert, so replication directly enlarges the low cost region of the collaboration domain. Third, after replication the residency is globally re-optimized, so $\mathbf{X}$ and $\mathbf{Y}$ are not frozen independently. The result is a placement whose candidate sets, GPU residency, and precisions are jointly tuned to the online routing objective, which is exactly the sense in which HetRoute performs offline-online co-design.

We solve this offline problem by a three-stage heuristic algorithm, summarized in Algorithm 1. We emphasize that, since P0 couples offline placement with causal online routing under time varying state, Algorithm 1 is a tractable approximation of the offline projection of P0, not an exact joint optimizer.

*Stage 1: Routing Probability Weighted GPU Residency*

Let $\pi_E$ denote the activation frequency of expert $E$ estimated on $\mathcal{T}_{\text{cal}}$, and let $p_{E,n}^{\text{route}} \in [0,1]$ denote the empirical probability that, given $E$ is activated, the online router selects the replica on $s_n$. The latter is obtained by simulating Algorithm 2 on $\mathcal{T}_{\text{cal}}$ under the current placement. The GPU residency benefit of promoting replica $(E,n)$ from CPU to GPU is the expected offload latency it removes, weighted by how often it is actually used:

$$g_{E,n} = \pi_E \cdot p_{E,n}^{\text{route}} \cdot \frac{v_{E,n}}{\beta_n} \tag{13}$$

Stage 1 greedily promotes replicas in decreasing order of $g_{E,n}$ subject to the GPU memory budget (c.1). The weighting by $p_{E,n}^{\text{route}}$ is what prevents a high frequency expert sitting behind a slow link or a weak GPU from occupying scarce GPU memory it would rarely benefit from, since such a replica has a small online selection probability.

*Stage 2: Unified Cost Driven Redundant Replication*

Stage 2 adds redundant replicas to enlarge the collaboration domain where it matters most. For a target expert $E$ currently served at unified cost $C_{i,l}^{\text{old}}(E) = \min_{(\hat{E},m)\in\mathcal{S}_E} C_{i,l}(E,\hat{E},m)$, adding a new replica on server $n$ (at a chosen precision $c_{E,n}$ and residency $y_{E,n}$) lowers the achievable cost to $C_{i,l}^{\text{new}}(E,n)$. The redundancy benefit is the expected reduction over the calibration traffic, net of a memory price:

$$b_{E,n} = \mathbb{E}_{(i,l)\sim\mathcal{T}_{\text{cal}}}\cdot[\, C_{i,l}^{\text{old}}(E) - C_{i,l}^{\text{new}}(E,n) \,] - \mu\, v_{E,n} \tag{14}$$

Importantly, because $C_{i,l}(\cdot)$ in (11) already contains the transmission, offload, computing queue, and quality terms, the replication decision is driven by the identical four factor cost used online. This closes the offline-online loop: a replica is created precisely when it is expected to reduce the online routing cost, and the precision $c_{E,n}$ is chosen by the same trade-off, so a low precision replica with small footprint may be preferred when its quality term $q_{E,n}$ stays within budget but its small $v_{E,n}$ enables GPU residency. Stage 2 repeatedly inserts the replica with the largest positive $b_{E,n}$ until no positive benefit insertion fits the memory budgets (c.1) to (c.2) or a replication cap is reached.

*Stage 3: Residency Re-Optimization*

After new replicas are added, the GPU residency assignment of Stage 1 is re-run globally over all replicas (old and new), with a diminishing returns guard: a second GPU resident copy of the same expert on a different server is admitted only if its marginal $g_{E,n}$ exceeds that of the best not-yet-resident distinct expert. This prevents wasting GPU memory on redundant GPU copies of one expert while a high frequency, CPU-only distinct expert starves, and it ensures $\mathbf{X}$ and $\mathbf{Y}$ are co-adapted rather than fixed in isolation.

Table I

Alaogirtim1: Routing-cost-coupled offline expert placement

**Input:** baseline expert set $\{E_l\}$; calibration traffic $\mathcal{T}_{\text{cal}}$; memory budgets $\{G_n^M, C_n\}$; network/compute params $\{B_{\text{link}}^{(\text{m,n})}, \beta_n, F_n\}$; precision set $Q$; replication cap $K_{\text{rep}}$; memory price $\mu$.

**Output:** deployment $\mathbf{X}$, residency $\mathbf{Y}$, precisions $\{c_{E,n}\}$, exact candidate sets $\{\mathcal{C}_E^{ex}\}$.

**1:** Initialize $\mathbf{X}$ with the baseline deployment satisfying the full-precision invariant (c.4) and set $\mathbf{Y} \leftarrow \mathbf{0}$.
**2:** Estimate activation frequencies $\{\pi_E\}$ on $\mathcal{T}_{\text{cal}}$.
// *Stage 1: routing-probability-weighted GPU residency*
**3:** Simulate Algorithm 2 on $\mathcal{T}_{\text{cal}}$ to obtain $\{p_{E,n}^{\text{route}}\}$.
**4:** Compute $g_{E,n}$ by (13) for all deployed replicas.
**5: while** GPU budget (c.1) allows and a non-resident replica has $g_{E,n} > 0$ **do**
**6:** Promote the replica with the largest $g_{E,n}$: $y_{E,n} \leftarrow 1$.
**7: end while**
// *Stage 2: unified-cost-driven redundant replication*
**8: for** $r = 1$ to $K_{\text{rep}}$ **do**
**9: for** each feasible $(E,n)$ with $x_{E,n} = 0$ and each $c_{E,n} \in Q$ **do**
**10:** Compute $b_{E,n}$ by (14) using the online cost (11').
**11: end for**
**12:** $(E^*, n^*, c^*) \leftarrow \arg\max b_{E,n}$.
**13: if** $b_{E^*,n^*} \le 0$ or memory (c.1) to (c.2) infeasible **then break**.
**14:** Add replica: $x_{E^*,n^*} \leftarrow 1$, $c_{E^*,n^*} \leftarrow c^*$.
**15: end for**
// *Stage 3: global residency re-optimization*
**16:** Recompute $g_{E,n}$ over all replicas; re-assign $\mathbf{Y}$ greedily under (c.1) with the diminishing-returns guard.
**17:** Compute $\mathcal{C}_E^{ex} = \{(E,n) \mid x_{E,n} = 1\}$ for all $E$.
**18: return X**, **Y**, $\{c_{E,n}\}$, $\{\mathcal{C}_E^{ex}\}$

The algorithm contains three parts.

The first part (**lines 1-7**) establishes a routing aware GPU

residency. It begins from a baseline deployment that already satisfies the full precision invariant (c.4), so that every expert is reachable at full precision and the online fallback can never be quality infeasible. It then profiles the activation frequencies on the calibration traffic and, instead of promoting experts by raw frequency, it first simulates the online router (Algorithm 2) to learn how often each replica would actually be chosen. The residency benefit $g_{E,n}$ in (13) multiplies the offload latency saved by this selection probability, so GPU memory is spent on the replicas that the online stage will genuinely rely on. Replicas are promoted greedily until the GPU budget (c.1) is exhausted, which yields the residency map $\mathbf{Y}$ that maximizes the expected online offload saving under a fixed deployment.

The second part (**lines 8-15**) enlarges the candidate collaboration domain through unified cost driven replication. For every feasible server and every admissible precision, it evaluates the redundancy benefit $b_{E,n}$ in (14), which is the expected reduction of the online unified cost (11) minus a memory price. Because this benefit is computed from the same four factor cost the router optimizes online, replication is steered toward the experts and servers where the online routing cost is currently highest. For example, a hot expert whose only replica sits behind a congested link, or a quality sensitive expert that would otherwise be served by a lossy substitute. The precision of each new replica is selected jointly: a compact low-precision replica is preferred when it both fits the GPU and keeps its quality term within budget, whereas a full precision replica is added when quality dominates. Replication stops when no insertion yields a positive benefit or the memory and replication-cap limits are reached.

The third part (lines **16-18**) couples deployment and residency rather than freezing them separately. After new replicas change the memory landscape, the residency is re-optimized globally over the enlarged replica set, with a diminishing-returns guard that forbids a second GPU copy of an expert from displacing a distinct, more valuable expert. The procedure returns the deployment, residency, and precision triple $(\mathbf{X}, \mathbf{Y}, \{c_{E,n}\})$ that jointly minimizes the expected online routing cost, which Algorithm 2 then consumes as its candidate structure.

*Offline complexity*. Stage 1 costs $O(|\mathcal{E}|N)$ to score and sort residency candidates; Stage 2 costs $O(K_{\text{rep}}|\mathcal{E}|N|\mathcal{Q}|)$ over the calibration traffic; Stage 3 repeats the Stage-1 scan. As this runs once per epoch rather than per token, its cost is amortized and does not affect online latency.

## V. Communication Aware Online Skip Exit Server-expert Selection

### A. Online Decision State and Per-Step Cost

The online stage processes tokens layer by layer over the placement produced by Algorithm 1. For token $\tau_i$, let $\theta_{i,l}$ be the server on which the token representation resides before the $l$-th MoE layer, initialized to the home server $\theta_{i,l} = a_i$. The token then follows the trajectory of its executed experts: after a layer is computed on the participating set $\mathcal{P}_{i,l}$, its partial outputs are gathered to the aggregation server $\theta_{i,l+1}$ chosen by (9), which becomes the residing server of the next layer. Consecutive layers executed on the same server therefore incur no transmission, and the token returns to $a_i$ only to emit the final output.

At layer $l$, the gating network produces the Top-$k$ activated set $\mathcal{K}_{i,l}$. For each target expert $E \in \mathcal{K}_{i,l}$, the router constructs candidates according to the exact-first policy defined in Section III.B. It first forms the exact candidate set $\mathcal{C}_E^{ex}$ from all deployed replicas of $E$ and applies the quality guard and the normal stability guard to these candidates. If at least one exact candidate remains admissible, the router confines its selection to $\mathcal{C}_E^{ex}$. If no exact candidate remains admissible, the router activates the substitute candidate set $\mathcal{C}_E^{sub}$ and applies the same two guards to the substitute candidates. This rule ensures that substitute execution is never mixed with exact execution: it is used only when exact execution cannot provide an admissible assignment under the current online state. Under this policy, the target $E$ is always the expert selected by the gating network, while the executed replica $\hat{E}$ is selected by the router. The quality increment of an assignment $(\hat{E}, n)$ serving target $E$ is $\Delta\Lambda(E, \hat{E}, n) = q_{\hat{E},n} + Q_{\text{sub}}(E, \hat{E})$ , where $Q_{\text{sub}}(E, \hat{E}) = 0$ and $q_{E,n} = 0$ at full precision, as defined in Section III.C.

The cost of a single candidate assignment is the unified per-assignment cost $C_{i,l}(E, \hat{E}, n)$ in (11), measured from the current residing server $\theta_{i,l}$.

Two online guards prune candidates during the normal routing path. The first is the hard per-token quality guard:

$$\Lambda_i^{\text{acc}} + q_{\hat{E},n} + Q_{\text{sub}}(E, \hat{E}) \leq \Lambda_i^{\max} \tag{18}$$

which is never relaxed under any circumstances. The second is the normal per-server stability guard:

$$W_n(\mathbf{X}, \mathbf{Z}; t) + w_{\hat{E}} \leq F_n \, \Delta t \tag{19}$$

which prevents the router from admitting a candidate that would push server $s_n$ beyond its window capacity. Any candidate violating (18) or (19) is removed from the normal search space.

After applying both guards, if no admissible assignment can be formed for some target expert, either from $\mathcal{C}_E^{ex}$ or $\mathcal{C}_E^{sub}$, Algorithm 2 invokes the emergency fallback. In fallback mode, each such target is routed to the full precision exact replica guaranteed by (c.4). The quality guard (18) remains mandatory and is satisfied with a zero increment (see Property 1). The stability guard (19) is relaxed: the fallback assignment is admitted into the server queue, and the resulting delay is reflected by the $Q_n(t)/F_n$ queueing term in the unified cost model. Therefore, fallback may increase latency but never violates the quality budget.

### B. Collaboration Set Selection

The distinguishing feature of HetRoute is that a layer is routed as a set, not as $k$ independent choices. Because the experts run in parallel and the layer latency is the bottleneck cost $D_{i,l}$ in (8), two targets placed on the same server share one fan-out branch but compete on its compute branch, whereas two targets on different servers add a parallel branch but enlarge fan-out and fan-in. These interactions are invisible to per-expert greedy selection. The router therefore solves the per-layer subproblem:

$$\mathbf{P}_{i,l}: \min_{\boldsymbol{z}} D_{i,l}(\boldsymbol{z})$$
$$s.t. \text{ (c.6), (c.7), (18), (19)}$$

over the candidate collaboration domain $\mathcal{C}_{i,l}$ in (5). When the domain is small, i.e., $|\mathcal{C}_{i,l}| \leq \bar{R}^k$, the subproblem is solved by exact enumeration of all feasible complete assignments;

otherwise, a beam search of width $B$ is used, expanded by the negative marginal cost:

$$\varphi_{i,l}\big(E,\hat{E},n \mid \mathbf{z}^{(t)}\big) = -\big[D_{i,l}\big(\mathbf{z}^{(t)} \cup \{z_{i,l,E,\hat{E},n} = 1\}\big) - D_{i,l}(\mathbf{z}^{(t)})\big] \quad (20)$$

and refined by 1-exchange local search. We stress that $\varphi$ is only an online computable expansion surrogate for the set objective; the acceptance of any partial solution re-evaluates the full bottleneck cost $D_{i,l}$ in (8), so the gather coupling to $\theta_{i,l+1}$ that $\varphi$ cannot capture is always handled exactly at the set level.

The adaptive unit costs $\{\alpha_j\}$ in the cost model are refined online by a projected feedback rule. Let $\rho_j(t)$ be the realized average contribution of cost component $j$ (transmission, offload, compute-queue, quality) over the recent window, and $\bar{\rho}_j$ its target share. Then:

$$\alpha_j \leftarrow \frac{[\alpha_j+\eta(\rho_j(t)-\bar{\rho}_j)]_+}{\sum_{j'}[\alpha_{j'}+\eta(\rho_{j'}(t)-\bar{\rho}_{j'})]} \quad (21)$$

which raises the price of a component that is becoming the system bottleneck and renormalizes the weights to the simplex. This is the rule referenced as the feedback step in Algorithm 2.

Table II

Algorithm2: Communication aware online collaborative Top-k routing

**Input**: $\mathbf{X},\mathbf{Y},\{c_{E,n}\},\{\mathcal{C}_E^{ex}\}$ from Algorithm 1; substitute sets $\{\mathcal{V}_{l,r_l(i)}\}$; unit costs $\{\alpha_j\}$; budgets $\{\Lambda_i^{\max}\}$; params $\{B_{\text{link}}^{(m,n)},\beta_n,F_n\}$; slot $\Delta t$; beam width $B$; local threshold $\tau_{\text{loc}}$.
**Output**: routing strategy $\mathbf{Z}$, token trajectories.
**1: for** each token $\tau_i$: set $\theta_{i,1} \leftarrow a_i$, $\Lambda_i^{\text{acc}} \leftarrow 0$.
**2: for** layer $l = 1$ to $L$ **do**
**3: for** each token $\tau_i$ **do**
**4:** $\mathcal{K}_{i,l} \leftarrow$ Top-$k$ activated experts (gating network).
*// Exact-first candidate construction*
**5: for** each $E \in \mathcal{K}_{i,l}$ **do**
**6:** $\mathcal{C}_E^{ex} \leftarrow \{(E,n) \mid x_{E,n} = 1\}$
**7:** Prune $(E,n) \in \mathcal{C}_E^{ex}$ violating guard (18) or (19)
**8: if** $\mathcal{C}_E^{ex} \neq \emptyset$ **then**
**9:** $\mathcal{C}_E \leftarrow \mathcal{C}_E^{ex}$ *(exact path; substitutes NOT activated)*
**10: else**
**11:** $\mathcal{C}_E^{sub} \leftarrow \{(\hat{E},n) \mid \hat{E} \in \mathcal{V}_{l,r_l(i)}, \hat{E} \neq E, x_{\hat{E},n} = 1\}$
**12:** Prune $(\hat{E},n) \in \mathcal{C}_E^{sub}$ violating guard (18) or (19).
**13:** $\mathcal{C}_E \leftarrow \mathcal{C}_E^{sub}$
**14: end if**
**15: end for**
**16:** $\mathcal{C}_{i,l} \leftarrow \bigcup_{E\in\mathcal{K}_{i,l}} \mathcal{C}_E$ *(active collaboration domain)*
*// Dominance fast path (not unconditional local-first)*
**17: if** all $E$ local & GPU-resident on $\theta_{i,l}$, and $Q_{\theta_{i,l}}(t)/F_{\theta_{i,l}} \leq \tau_{\text{loc}}$, and $D_{\text{local}} \leq$ best remote branch lower bound **then**
**18:** assign all $E$ to $\theta_{i,l}$; $\theta_{i,l+1} \leftarrow \theta_{i,l}$; **goto 26**.
*// Collaboration-set selection*
**19: if** $|\mathcal{C}_{i,l}| \leq \bar{R}^k$ **then**
**20:** $\mathbf{Z}_{i,l} \leftarrow \arg\min_{\text{feasible}} D_{i,l}(\mathrm{z})$ *(exact enumeration)*
**21: else**
**22:** $\mathbf{Z}_{i,l} \leftarrow BeamSearch(\mathcal{C}_{i,l}, D_{i,l}, B)$ then 1-exchange refine
**23: end if**
*// Emergency fallback (quality-safe, stability-relaxed)*
**24: if** no feasible complete assignment **then**
**25: for** each $E \in \mathcal{K}_{i,l}$ with no admissible candidate **do**
**26:** Assign $E$ to full-precision exact replica $n_0$ (invariant c.4); admit into queue of $s_{n_0}$.
**27: end for**
**28: end if**
**29:** $\mathcal{P}_{i,l} \leftarrow$ servers in $\mathbf{Z}_{i,l}$; $\theta_{i,l+1} \leftarrow$ by (9).
**26:** Scatter $\theta_{i,l} \rightarrow \mathcal{P}_{i,l}$, execute, gather $\rightarrow \theta_{i,l+1}$.
**27:** $\Lambda_i^{\text{acc}} \mathrel{+}= \sum_{E\in\mathcal{K}_{i,l}}(q_{\hat{E},n} + Q_{\text{sub}}(E,\hat{E}))$; update backlog $Q_n(t)$.
**28: end for**
**29:** Refine $\{\alpha_j\}$ by feedback rule (21).
**30: end for**
**31: return Z**, trajectories.

The algorithm contains three parts.

The first part (**lines 1-16**) builds the candidate collaboration domain for each token-layer pair. Every token starts on its home server with zero accumulated degradation. At a layer, the gating network selects the $k$ activated targets.

The router first collects the exact candidates $\mathcal{C}_E^{ex}$ formed by the deployed replicas of each target expert $E$. These candidates are pruned by the quality guard (18) and the normal stability guard (19). Only when no exact candidate remains admissible does the router activate the substitute candidate set $\mathcal{C}_E^{sub}$ from the similarity group $\mathcal{V}_E$ and subject them to the same guards. Hence, substitution is triggered not merely by the physical absence of an exact replica, but by the absence of an admissible exact assignment under the current quality, load, and reachability state. This keeps the routing policy exact-first while still allowing the system to continue serving tokens when exact execution is temporarily infeasible or excessively constrained. The union of the surviving candidates forms the collaboration domain $\mathcal{C}_{i,l}$, whose size may exceed $k$ because of redundancy, while the eventual participating set will contain at most $k$ servers.

The second part (**lines 15-18**) performs the set-level routing that constitutes the core of the method. Before the general search, a dominance fast path is checked: if all activated experts have a GPU-resident replica on the current residing server, the local queue is light, and local execution provably dominates the best remote branch, the layer is executed in place at zero transmission. This is deliberately not an unconditional local-first rule; when the local GPU is contended or slower than a remote GPU resident replica behind a fast link, the local option is treated as one ordinary candidate in $\mathcal{C}_{i,l}$ and competes on the unified cost like any other. Otherwise, the router solves the per-layer subproblem $\mathrm{P}_{i,l}$ by minimizing the bottleneck cost (8) over feasible complete assignments: exactly by enumeration when the domain is small, and by a marginal-cost beam search with 1-exchange refinement when it is large. Because the objective is the bottleneck cost rather than a sum, the search captures the fan-out and fan-in interactions among co-selected servers that per-expert greedy selection misses. If, after applying the normal guards to both $\mathcal{C}_E^{ex}$ and $\mathcal{C}_E^{sub}$, no admissible complete assignment can be formed for the current token-layer pair, Algorithm 2 invokes the emergency fallback. For each affected target expert $E$, the fallback selects the full precision exact replica guaranteed by constraint (c.4). Since this replica has $q_{E,n} = 0$ and $Q_{\text{sub}}(E,E) = 0$, the assignment adds zero to the accumulated quality degradation and therefore cannot violate the hard quality budget (18). If the selected server is already saturated within the current scheduling window, the assignment is admitted into its queue rather than rejected; the overload is reflected as additional queueing delay in the unified cost model. This design makes Algorithm 2 complete and quality safe: fallback is permitted to pay latency, but it is never permitted to pay quality loss. Finally, the participating set and the next residing server are fixed, the latter by the gather-minimizing rule (9).

The third part (**lines 29-31**) executes and adapts. The token is scattered from $\theta_{i,l}$ to the participating servers, the experts run in parallel, and the partial outputs are gathered to $\theta_{i,l+1}$, which becomes the residing server of the next layer. The accumulated degradation and the per-server backlog are updated, and the unit costs $\{\alpha_j\}$ are refined by the projected-feedback rule (21) so that whichever cost component is becoming the system bottleneck is penalized more heavily in subsequent decisions. The procedure repeats across layers and tokens and returns the complete routing strategy together with the token trajectories.

*C. Theoretical Properties*

In this section, we investigate the properties of feasibility, structural, optimality, and complexity guarantees of HetRoute.

**Property 1 (Fallback feasibility).** *Under the full precision constraint (c.4) and the queueing relaxed stability model, Algorithm 2 always terminates with a complete and feasible routing* $\mathbf{Z}$*: every activated target is assigned, and no assignment violates the quality budget (18).*

*Proof.* Fix any token $\tau_i$, layer $l$, and target $E \in \mathcal{K}_{i,l}$. By (c.4) there exists a server $n_0$ with $x_{E,n_0} = 1$ and $c_{E,n_0} = c_{\max}$, i.e., a full precision exact replica. Hence the candidate set is non-empty, $(E, n_0) \in \mathcal{C}_E$, so the fallback in line 17 is always reachable. We verify it passes both guards.

1) Quality: For any activated target $E$, constraint (c.4) guarantees a server $n_0$ with $x_{E,n_0} = 1$ and $c_{E,n_0} = c_{\max}$. By the normalization in Section III.C, $q_{E,n_0} = 0$ and $Q_{\mathrm{sub}}(E, E) = 0$. Hence, the quality increment of the fallback assignment is $\Delta\Lambda(E, E, n_0) = q_{E,n_0} + Q_{\mathrm{sub}}(E, E) = 0$ . Since $\Lambda_i^{\mathrm{acc}}$ is initialized to zero at the start of each token and only non-negative increments are added during normal routing, the fallback assignment leaves $\Lambda_i^{\mathrm{acc}}$ unchanged and $\Lambda_i^{\mathrm{acc}} \leq \Lambda_i^{\max}$ is preserved. The quality guard (18) is satisfied unconditionally.

2) Stability: During normal routing, candidates violating the stability guard (19) are pruned. If this pruning leaves no complete feasible assignment for some target $E$, even after activating $\mathcal{C}_E^{sub}$, the fallback mechanism is triggered. The fallback assignment to $(E, n_0)$ is not rejected due to temporary overload. Instead, its workload $w_E$ is appended to the backlog of $s_{n_0}$, increasing the queueing term $Q_{n_0}(t)/F_{n_0}$ in the cost model and raising the realized inference latency. The assignment itself remains well defined. Thus, temporary violation of the normal stability guard (19) is converted into queueing latency rather than infeasibility. Since Quality guarantees a quality guard feasible assignment for every target, and since (c.7) requires each target to be assigned exactly once, iterating over all $E \in \mathcal{K}_{i,l}$ and all layers $l$ yields a complete routing $\mathbf{Z}$ for $\tau_i$. Algorithm 2 therefore always terminates with a complete, quality feasible routing.

Combining 1) to 2), every target obtains at least one guard feasible assignment, so by (c.7) the layer is completely routed; iterating over $l$ and $\tau_i$ yields a complete feasible $\mathbf{Z}$. ■

**Property 2 (Bounded participating servers).** *For every token-layer pair, i.e.,* $|\mathcal{P}_{i,l}| \leq k$*, regardless of the collaboration domain size* $|\mathcal{C}_{i,l}|$.

*Proof.* By the assignment constraint (c.7), each target $E \in \mathcal{K}_{i,l}$ has exactly one nonzero routing variable $z_{i,l,E,\hat{E},n} = 1$. The number of such assignments equals $|\mathcal{K}_{i,l}| = k$ . The participating set $\mathcal{P}_{i,l}$ in (6) is the set of distinct servers appearing in these $k$ assignments, hence $|\mathcal{P}_{i,l}| \leq k$, with equality when the $k$ targets are placed on $k$ distinct servers and $|\mathcal{P}_{i,l}| = 1$ when all are co-located. The collaboration domain $\mathcal{C}_{i,l} = \bigcup_E \mathcal{C}_E$ may contain more than $k$ servers because of redundant replicas, but only the selected servers enter $\mathcal{P}_{i,l}$, so the bound is independent of $|\mathcal{C}_{i,l}|$. This is consistent with constraint (c.8). ■

**Property 3 (Per-layer optimality for small $k$).** *If* $|\mathcal{C}_{i,l}| \leq \bar{R}^k$*, the exact enumeration in line 14 returns a global optimum of the per-layer subproblem* $P_{i,l}$ *under the current backlog state* $Q_n(t)$.

*Proof.* The feasible region of $P_{i,l}$ is the set of complete assignments that pick, for each target $E$, one candidate from its pruned set $\mathcal{C}_E$ (constraints (c.6) to (c.7)) such that the per-token quality guard (18) and the per-server stability guard (19) hold. This region is contained in the product $\prod_{E \in \mathcal{K}_{i,l}} \mathcal{C}_E$, whose cardinality is $\prod_E |\mathcal{C}_E| \leq \bar{R}^k$. The enumeration iterates over all complete assignments in this product, discards those violating (18) or (19), evaluates the bottleneck objective $D_{i,l}$ in (8) exactly for each survivor, using the current backlog $Q_n(t)$, so the queueing term is the true online value, and returns the minimizer. Because the search is exhaustive over the entire feasible region and the objective is evaluated without approximation, the returned assignment attains $\min_{\mathbf{z}} D_{i,l}(\mathbf{z})$ over the feasible set, i.e., it is globally optimal for $\mathrm{P}_{i,l}$. We claim per-layer, not cross-layer, optimality: the accumulated budget $\Lambda_i^{\mathrm{acc}}$ and the backlog $Q_n(t)$ couple successive layers and tokens, so the layer-wise optima need not compose into a global optimum of P0. ■

**Property 4 (Local optimality of the exchange refinement).** *For* $|\mathcal{C}_{i,l}| > \bar{R}^k$*, the beam search followed by 1-exchange refinement terminates at a solution* $\mathbf{z}^*$ *that is locally optimal with respect to single-target reassignment: no feasible reassignment of one target to a different candidate strictly decreases* $D_{i,l}$.

*Proof.* The refinement accepts a single-target reassignment only if it is feasible (preserves (18) to (19)) and strictly decreases the bottleneck objective $D_{i,l}$. Each accepted move thus produces a strictly smaller objective value. The feasible set is finite (at most $\bar{R}^k$ complete assignments) and $D_{i,l} \geq 0$ is bounded below, so a strictly decreasing sequence of distinct objective values cannot be infinite; the refinement therefore terminates after finitely many moves. At termination, by the acceptance rule, there is no feasible single-target reassignment that strictly decreases $D_{i,l}$, which is precisely local optimality over the 1-exchange neighborhood. We do not claim a global approximation ratio: the bottleneck (min-max) structure together with the coupling guards (18) to (19) does not in general admit a constant-factor bound, and we instead report the empirical gap to an offline oracle in Section VI. ■

**Remark (recovering the additive special case).** When the layer reduces to a single activated server ($|\mathcal{P}_{i,l}| = 1$) or the targets do not interact on any branch, the bottleneck objective (8) degenerates into the separable cost $\sum_E \min_{(\hat{E},n)} C_{i,l}(E, \hat{E}, n)$, and the set selection reduces to the per-target rule: $\arg\max_{(\hat{E},n)} \varphi_{i,l}$, i.e., $\max \varphi \equiv \min C_{i,l}$. Thus, the per-expert greedy choice is an exact special case of the collaboration set selection, and HetRoute strictly generalizes it by accounting for the fan-

out/fan-in coupling whenever the participating servers interact.

**Property 5 (Online complexity).** *The per-token routing cost is $O(Lk\bar{R}B)$ for the beam-search variant and $O(L\bar{R}^k)$ for the exact variant, scaling with the average replica count $\bar{R}$ rather than the server count $N$.*

*Proof.* Per layer, building the collaboration domain scans, for each of the $k$ targets, its at most $\bar{R}$ candidates and applies the $O(1)$ guards, costing $O(k\bar{R})$. In the beam-search variant, the search maintains $B$ partial sets. At each of the $k$ expansion rounds, extends every partial set over at most $\bar{R}$ candidates. Each extension evaluates the marginal cost (20) in $O(1)$ using incrementally maintained branch maxima, giving $O(k\bar{R}B)$ per layer, and a constant number of 1-exchange sweeps adds the same order. Over $L$ layers this is $O(Lk\bar{R}B)$. In the exact variant, there are at most $\bar{R}^k$ complete assignments, each evaluated in $O(k)$ for the bottleneck objective, giving $O(\bar{R}^k)$ per layer (constant for small $k$) and $O(L\bar{R}^k)$ in total. Since Algorithm 1 bounds each expert's replicas by $\bar{R} \ll N$, the candidate sets are small and the online overhead is independent of the number of edge servers $N$, which makes token-level routing practical for typical Top-$k$ MoE where $k \in \{2,4,8\}$. ■

## VI. Performance Evaluation

### A. Experimental Setup

**Experimental setup**. HetRoute is implemented in PyTorch and evaluated on a trace-driven heterogeneous edge MoE inference testbed. The edge system consists of 10 geo-distributed edge servers connected by ordinary Internet links rather than a dedicated cluster fabric. The servers are heterogeneous in both computation and memory: their GPU computing capacities span 18 to 120 TFLOPS, their GPU memory capacities span 12 to 48 GB, and their CPU memory capacities span 96 to 512 GB. The GPU-CPU transfer bandwidths range from 12 to 64 GB/s, covering PCIe- and NVLink-like memory-transfer conditions. The inter-server available bandwidths range from 0.8 to 10 Gbps, and the propagation-plus-queuing latencies range from 0.3 to 22 ms. Each user is attached to one access edge server following a non-uniform spatial distribution, and requests arrive at a default rate of 40 requests per second unless stated otherwise. The queueing backlog of each server is updated at the scheduling window granularity, and the online router observes the current link bandwidths, GPU backlogs, and candidate expert locations before making the token-layer routing decision.

**Models, tasks, and calibration**. We evaluate HetRoute on three MoE models of increasing scale, namely Switch-Base-8E, Qwen-MoE-A 2.7B, and Mixtral-8x7B. The Top-k value is set to 2 for Switch-Base-8E and Mixtral-8x7B, and follows the default sparse routing configuration for Qwen-MoE-A 2.7B. Inference quality is measured on WikiText-103 by perplexity, SQuAD by F1 score, and GSM8K by accuracy. All offline quantities required by Algorithm 1, including expert activation frequencies, online replica selection probabilities, per-replica quantization losses, substitution losses, and unified-cost-driven replication benefits, are computed once on a held-out calibration set with the backbone strictly frozen. Unless otherwise stated, the per-token quality degradation budget is set to 2%, the beam width of the online set-level router is W=8, the exact enumeration threshold is R=64 candidate assignments, and the memory budget ratio of redundant expert deployment is 2.0.

**Baselines**. We compare HetRoute with four representative methods that cover the most relevant design space. 1) Prism []; 2) EdgeShard []; 3) MoE-Infinity []; 4) Petals []. We further include three internal variants for the ablation study: HetRoute without set-level routing, HetRoute without GPU-CPU residency optimization, and HetRoute without quality-aware precision selection. The metrics are the average inference latency, the tail latency, the cross-server communication volume, the remote execution ratio, the CPU-offload ratio, the throughput, and the inference quality.

### B. Overall Performance Comparison

We first compare the overall performance on Mixtral-8x7B, as shown in Fig. 2, which reports the average latency, the P99 latency, the cross-server traffic per one thousand tokens, the remote execution ratio, the CPU-offload ratio, and the throughput.

As shown in Fig. 2(a), HetRoute attains the lowest average latency of 156 ms, which is 59.0% lower than EdgeShard, 49.5% lower than Petals, 38.6% lower than MoE-Infinity, and 28.1% lower than Prism. The gain over EdgeShard and Petals comes from the fact that HetRoute routes activated experts rather than dense model partitions, thereby avoiding unnecessary remote layer/block traversal. The gain over MoE-Infinity comes from cross-server collaboration: MoE-Infinity reduces GPU-CPU offloading on one serving node, whereas HetRoute can select a remote GPU-resident replica when the local replica is CPU-resident or queued. The gain over Prism comes from the online set-level router, which evaluates latency complete Top-k collaboration set instead of optimizing placement alone. Fig. 2(b) shows that HetRoute also reduces the P99 latency to 286 ms, 58.0% lower than EdgeShard and 33.2% lower than Prism. This tail reduction is important because the worst cases in distributed MoE serving are usually caused by a combination of weak links, CPU-resident experts, and unlucky multi-hop token trajectories.

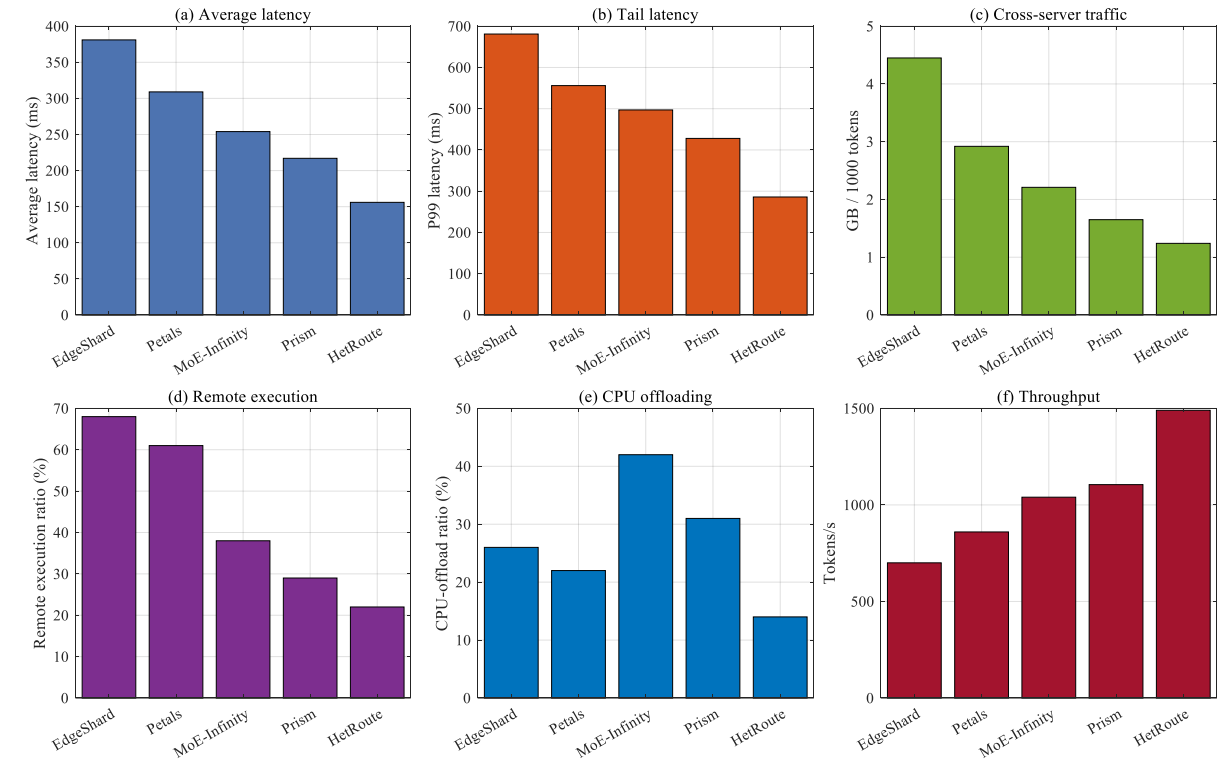

Fig. 2 Overall performance

Fig. 2(c) shows that the cross-server traffic of HetRoute is only 1.24 GB per one thousand tokens, 72.1% below EdgeShard, 57.5% below Petals, and 24.8% below Prism. Fig. 2(d) further shows that the remote execution ratio of HetRoute is 22%, whereas EdgeShard and Petals incur 68% and 61%, respectively. The reduction does not mean that HetRoute always chooses local execution; rather, it chooses remote execution only when the remote branch is sufficiently beneficial after considering transmission, queueing, offloading, and quality. Consistently, Fig. 2(e) shows that the CPU-offload ratio is reduced to 14%, compared with 42% for MoE-Infinity and 31% for Prism. The reason is that Algorithm 1 places frequently and actually selected replicas in GPU memory, while Algorithm 2 avoids CPU-resident candidates whenever a cheaper GPU-resident collaboration path exists. As a result, Fig. 2(f) shows that HetRoute achieves the highest throughput of 1490 tokens per second, which is 2.13 × that of EdgeShard and 1.35 × that of Prism.

### *C. Tail Latency and Collaborative Routing Behavior*

Fig. 3 reports the latency distribution and the expert execution-type breakdown. Fig. 3(a) plots the empirical CDF of the per-request latency, and Fig. 3(b) decomposes all MoE layer-level expert executions into local GPU exact, local CPU exact, remote GPU exact, remote CPU exact, substitute execution, and emergency fallback.

From Fig. 3(a), HetRoute shifts the whole latency distribution to the left and compresses the long tail. Its median latency is 134 ms and its P99 latency is 286 ms, while Prism reaches a median latency of 183 ms and a P99 latency of 428 ms. The difference becomes larger in the tail because HetRoute explicitly accounts for CPU-resident replicas and server backlog in the online set-level cost. A token is therefore not moved to a server that appears favorable by placement but becomes expensive after offloading and queueing are considered. From Fig. 3(b), 46% of HetRoute executions are local GPU exact executions and 24% are remote GPU exact executions, while only 9% and 5% are local CPU exact and remote CPU exact executions, respectively. Substitute execution accounts for 14% of the executions, and emergency fallback is only 2%. This confirms that substitute execution is a controlled last-resort mechanism rather than the main source of latency reduction. By contrast, MoE-Infinity has a much larger CPU-offload share, and EdgeShard has a much larger remote exact share. The breakdown verifies that HetRoute reduces latency by changing the execution composition toward GPU-resident exact replicas and by using substitute/fallback only when the exact normal path is constrained.

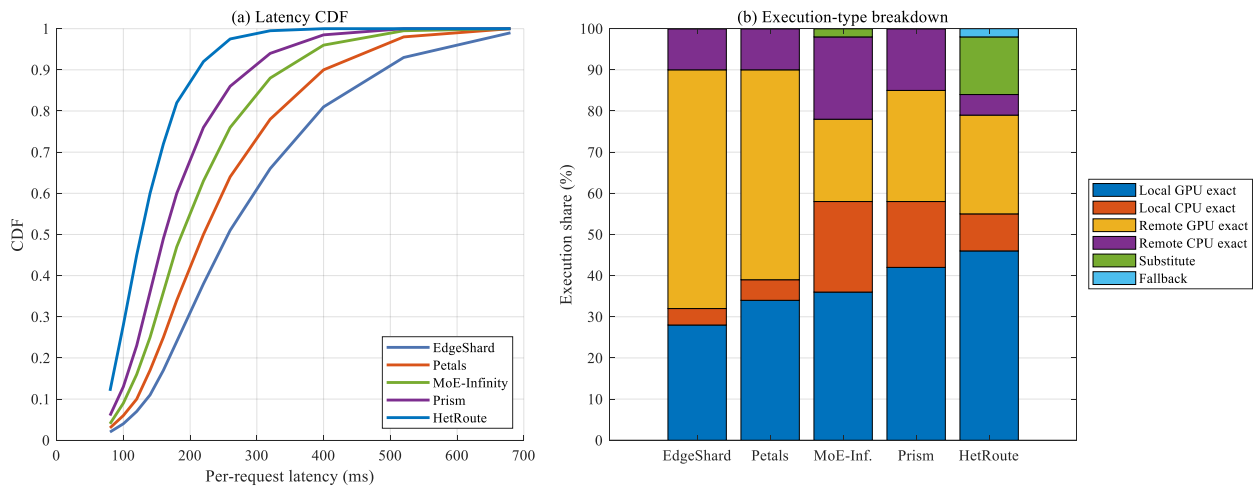

Fig. 3 Tail latency and collaborative routing behavior

### *D. Inference Quality and the Quality Budget*

Fig. 4 evaluates the inference quality. Fig. 4(a) reports the quality of representative methods on the three datasets, and Fig. 4(b) shows how the latency and the actual quality degradation of HetRoute vary with the per-token quality budget.

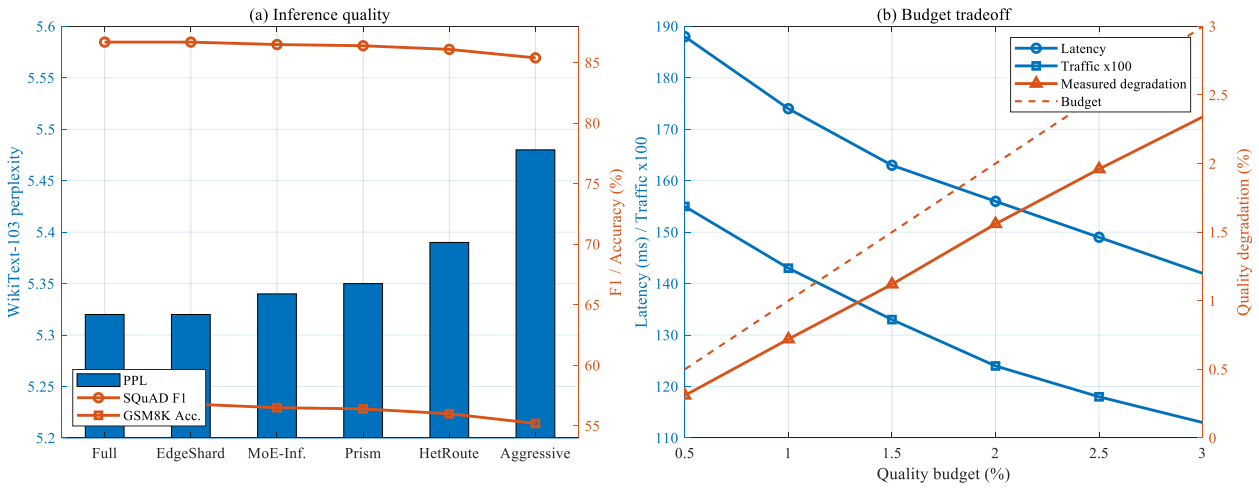

Fig. 4 Inference quality and quality budget

From Fig. 4(a), the quality of HetRoute remains close to the full-precision exact MoE reference. Compared with the centralized full precision model, the perplexity on WikiText-103 increases from 5.32 to 5.39, the SQuAD F1 drops from 86.7 to 86.1, and the GSM8K accuracy drops from 56.8% to 56.0%. This small and controlled loss is a direct consequence of the quality aware routing design: exact full precision replicas are always preserved by the offline invariant, quantized replicas are selected only when the profiled loss fits the per-token budget, and substitute execution is activated only when no admissible exact candidate remains. Prism keeps slightly higher quality because it uses fewer substitute or low-precision assignments, but it pays higher latency and traffic. MoE-Infinity also preserves quality well, but it cannot exploit cross-server GPU-resident alternatives when a hot expert is CPU-resident or queued.

From Fig. 4(b), as the quality budget is relaxed from 0.5% to 3.0%, HetRoute trades quality for latency in a smooth and monotone manner. The average latency decreases from 188 ms to 142 ms, while the measured degradation increases from 0.31% to 2.34%. In all cases, the measured degradation stays below the configured budget. The knee of the curve appears around the 2% budget, where the latency is already close to its minimum while the degradation remains below 1.6%. This justifies the default setting used in the other experiments.

### *E. Effect of Set-Level Collaborative Routing*

Fig. 5 examines the effect of set-level collaborative routing. Fig. 5(a) compares per-expert greedy routing and HetRoute set-level routing under different Top-k values, and Fig. 5(b) reports the distribution of participating servers per token-layer on Mixtral-8x7B.

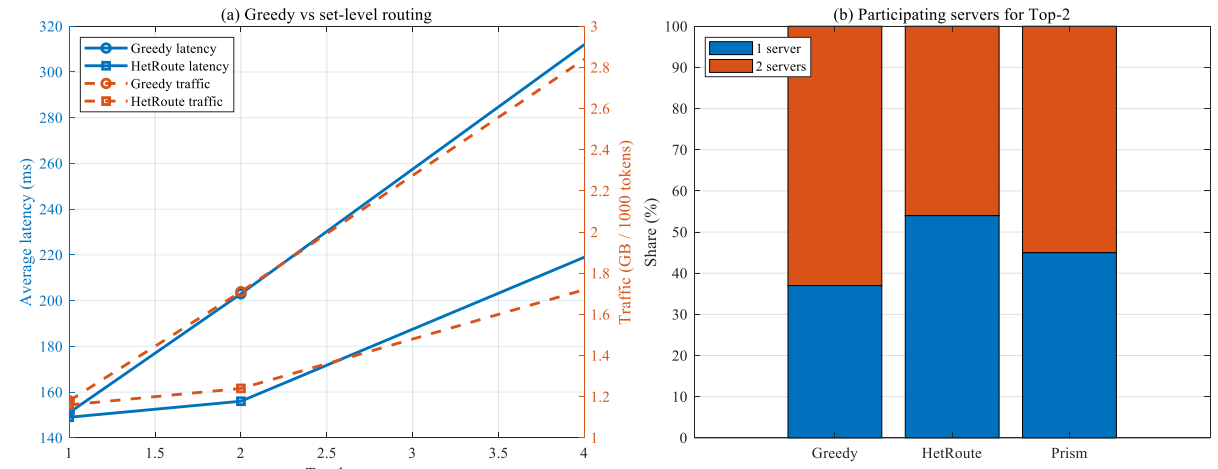

Fig. 5 Effect of set-level collaborative routing

As shown in Fig. 5(a), the advantage of set-level routing becomes more pronounced as Top-k increases. When k=1, the two strategies are almost identical, because the layer-level assignment degenerates into a single branch decision. When k=2, HetRoute reduces latency from 203 ms to 156 ms and cross-server traffic from 1.71 to 1.24 GB per one thousand tokens. When k=4, the latency reduction grows to 29.8%, because the fan-out and fan-in coupling among multiple activated experts becomes stronger. Per-expert greedy routing often selects the individually cheapest server for each expert, but the resulting set may involve too many remote branches. HetRoute instead evaluates the bottleneck layer cost of the complete collaboration set and therefore avoids a locally attractive but globally expensive combination.

Fig. 5(b) shows that HetRoute does not simply collapse all experts onto one server. With Top-2 routing, 54% of token-layer executions use one participating server and 46% use two participating servers. This indicates that HetRoute chooses co-location when it avoids communication, but still uses multi-server collaboration when parallel GPU-resident execution compensates for the additional fan-out and fan-in. This behavior is consistent with the design objective: the router is neither local-first nor remote-first, but cost-first under heterogeneous computation, communication, offloading, and quality constraints.

### *F. Effect of GPU-CPU Residency and Redundant Deployment*

Fig. 6 evaluates the effect of the memory budget ratio of redundant deployment, varied from 1.0 to 3.0. Fig. 6(a) reports the average latency, P99 latency, and CPU-offload ratio, and Fig. 6(b) reports the cross-server traffic and the remote GPU execution ratio.

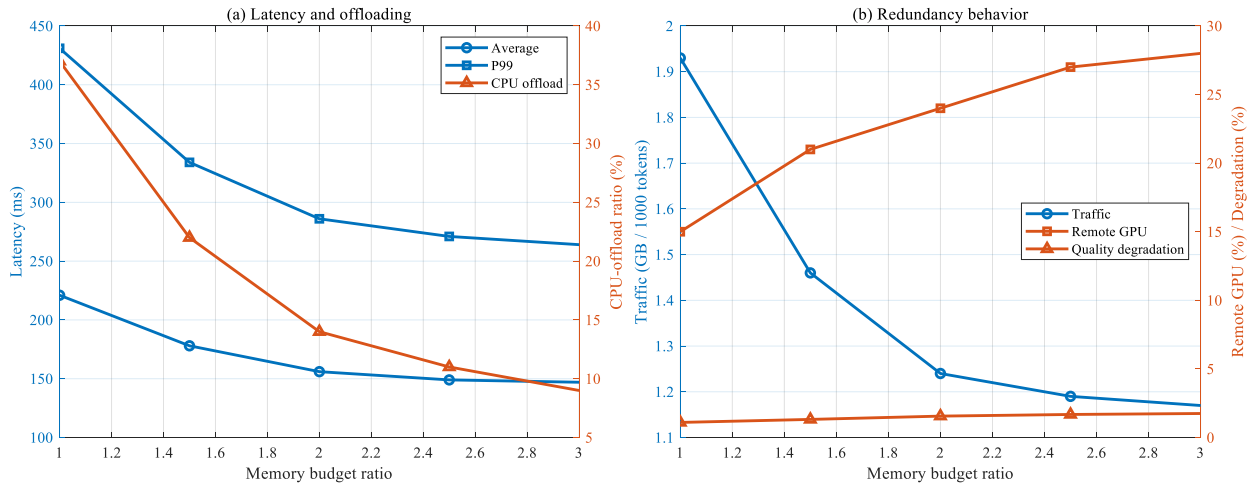

Fig. 6 Effect of GPU-CPU residency and redundant deployment

As shown in Fig. 6(a), increasing the memory budget ratio from 1.0 to 2.0 reduces the average latency from 221 ms to 156 ms and the P99 latency from 431 ms to 286 ms. The CPU-offload ratio decreases from 37% to 14%, because Algorithm 1 can promote more frequently selected replicas to GPU memory and create additional GPU-resident replicas on servers with favorable links. The improvement becomes smaller beyond a ratio of 2.0: increasing the ratio from 2.0 to 3.0 further reduces average latency by only 9 ms. This saturation occurs because most hot experts have already obtained at least one low-cost GPU-resident path.

Fig. 6(b) shows a non-monotone communication behavior. When the memory ratio increases from 1.0 to 2.0, cross-server traffic decreases from 1.93 to 1.24 GB per one thousand tokens, because more tokens can be served by nearby GPU-resident replicas. When the ratio is further increased, the remote GPU execution ratio rises slightly from 24% to 28%, but the traffic remains low. This is not a contradiction: additional replicas create more remote GPU-resident opportunities, and HetRoute uses them only when the saved offloading and queueing delay outweighs the extra transmission. Therefore, redundant deployment expands the collaboration domain without forcing unnecessary remote execution.

### *G. Comparison Across MoE Models*

Fig. 7 compares HetRoute with representative baselines on the three MoE models. Fig. 7(a) reports the average latency, Fig. 7(b) reports the cross-server traffic, and Fig. 7(c) reports the quality retention relative to the centralized full precision model.

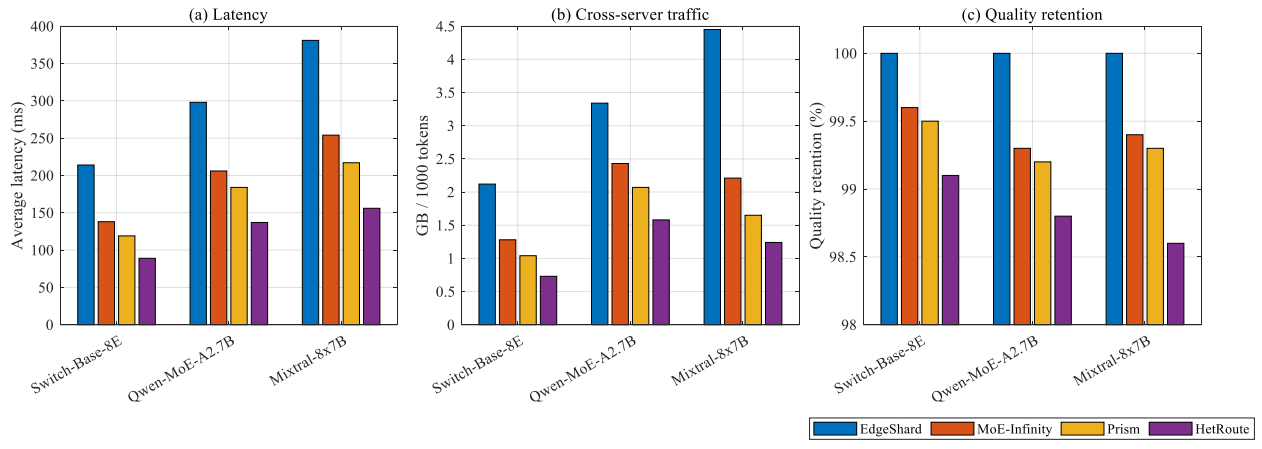

Fig. 7 Comparison across MoE models

As shown in Fig. 7(a), HetRoute consistently achieves the lowest latency on all three models. On Switch-Base-8E, HetRoute reduces average latency by 24.7% compared with Prism and by 35.5% compared with MoE-Infinity. On Mixtral-8x7B, the reduction grows to 28.1% and 38.6%, respectively. The absolute saving becomes larger as the model scales, because larger MoE models contain more layers and more experts, which increases both the probability that Top-k experts are distributed across servers and the opportunity for HetRoute to select a better collaboration set. Fig. 7(b) shows the same trend for communication: HetRoute reduces traffic from 2.07 to 1.58 GB on Qwen-MoE-A 2.7B and from 1.65 to 1.24 GB on Mixtral-8x7B compared with Prism. Fig. 7(c) shows that HetRoute retains at least 98.6% quality on all three models, confirming that the latency and communication savings are not obtained by uncontrolled quality degradation.

### *H. Ablation Study*

Fig. 8 isolates the contribution of each component on Mixtral-8x7B by removing set-level routing, GPU-CPU residency optimization, quality-aware precision selection, and redundant replication in turn. Fig. 8(a) reports latency and P99 latency, Fig. 8(b) reports cross-server traffic and CPU-offload ratio, and Fig. 8(c) reports the actual quality degradation.

From Fig. 8(a), removing any component increases latency. Disabling set-level routing causes the largest latency increase, from 156 ms to 203 ms, because the router degenerates to per-expert greedy selection and misses the fan-out/fan-in coupling among Top-k experts. Removing GPU-CPU residency optimization increases latency to 218 ms and P99 latency to 421 ms, which shows that ignoring CPU-resident experts leads to severe tail amplification. Removing quality-aware precision selection increases latency to 181 ms because the system must use larger full-precision replicas more often and has fewer memory efficient GPU-resident candidates. Removing redundant replication increases latency to 194 ms because the collaboration domain becomes too small to provide enough low-cost alternatives.

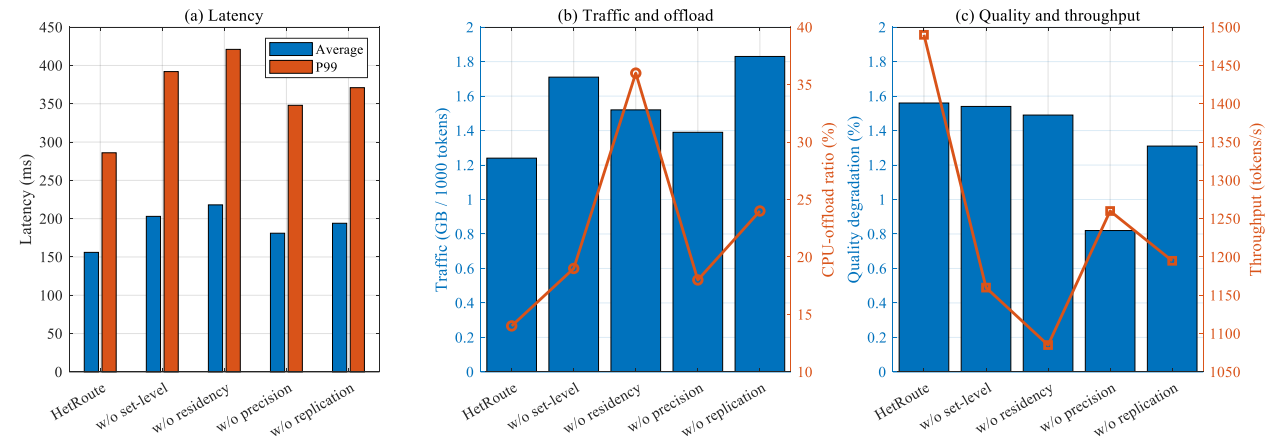

Fig. 8 Ablation study

Fig. 8(b) further confirms the mechanism. Without GPU-CPU residency optimization, the CPU-offload ratio rises from 14% to 36%. Without redundant replication, cross-server traffic increases from 1.24 to 1.83 GB per one thousand tokens because tokens are forced to travel to a smaller set of fixed expert locations. Fig. 8(c) shows that all variants satisfy the 2% quality budget, but the variant without quality-aware precision has the smallest degradation at the cost of higher latency. This indicates that the quality-aware precision module does not simply reduce quality. Instead, it uses the available quality budget to create compact GPU-resident replicas that reduce delay while remaining within the hard degradation bound.

### *I. Performance Under Different Request Loads*

Finally, Fig. 9 evaluates robustness to the request load, varied from 10 to 80 requests per second. Fig. 9(a) reports the average latency, Fig. 9(b) reports the P99 latency, and Fig. 9(c) reports the SLA satisfaction ratio, where the SLA target is 300 ms.

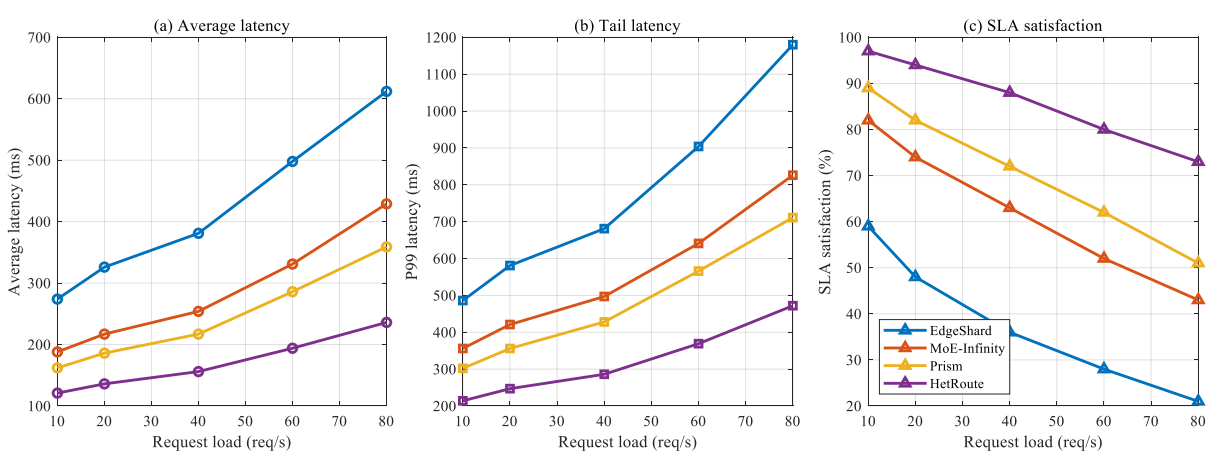

Fig. 9 Performance under different request loads

As shown in Fig. 9(a), the average latency of all methods grows with the request load because of increased GPU queueing and link contention. HetRoute grows the slowest: its latency increases from 121 ms at 10 requests per second to 236 ms at 80 requests per second, whereas Prism increases from 162 ms to 359 ms and EdgeShard increases from 274 ms to 612 ms. Fig. 9(b) shows that the same trend appears in tail latency. At 80 requests per second, HetRoute keeps the P99 latency at 472 ms, while Prism and EdgeShard reach 711 ms and 1180 ms, respectively. The reason is that HetRoute observes the current backlog in the unified cost and routes tokens away from temporarily congested servers when alternative GPU-resident candidates exist.

Fig. 9(c) shows that HetRoute still satisfies 73% of requests below the 300 ms target at 80 requests per second, whereas Prism, MoE-Infinity, and EdgeShard satisfy 51%, 43%, and 21%, respectively. This confirms that the proposed offline-online co-design is robust under heavy load: offline replication and GPU residency provide a rich candidate domain, while online set-level routing uses the current queue and link state to avoid the worst collaboration paths.

## VII. Conclusion

This paper presented HetRoute, a heterogeneous-cost-aware collaborative routing framework for distributed edge MoE inference. The core contribution is a unified per-assignment cost model that jointly captures cross-server transmission, GPU-CPU expert loading, heterogeneous GPU computation with queueing, and replica-level quantization quality loss. This model breaks the local-first assumption that underlies most existing distributed inference schemes: a remote GPU-resident hot expert may be faster than a local CPU-resident cold expert, and the optimal routing depends on the interplay of all four cost components rather than on any single factor. Building on this model, we designed a routing-cost-coupled offline deployment algorithm that determines server placement, GPU-CPU residency, and replica precision by estimating how each deployment decision affects the expected online routing cost. The online algorithm routes the Top-k activated expert set as a whole rather than making independent per-expert decisions, explicitly capturing the fan-out and fan-in coupling among co-selected servers. Theoretical analysis establishes the quality-safety of the emergency fallback, the participating-server bound, the per-layer optimality for small candidate domains, and the polynomial online complexity. Experiments on three MoE models over a heterogeneous 10-server edge testbed confirm that HetRoute consistently outperforms existing distributed inference and MoE serving baselines across latency, communication, and throughput metrics while maintaining controlled quality degradation. Future work includes extending HetRoute to handle time-varying expert activation distributions and incorporating network-topology-aware redundant replication strategies for dynamic edge environments..

## Acknowledgment

This work was supported in part by the grant from NSFC Grant no. 62571156, 62101159, 52475009, NSF of Shandong Grant no. ZR2021MF055, ZR2025QC666, the Research Grants Council of Hong Kong under the Areas of Excellence scheme grant AoE/E-601/22-R, and also the Opening Project of the Key Laboratory of Advanced Manufacturing and Intelligent Technology (Ministry of Education) at Harbin University of Science and Technology (KFKT202306).

Additionally, the authors used an AI-based language assistance tool to improve the clarity and readability of parts of the manuscript, particularly the Abstract and Introduction, and all technical content, analysis, and conclusions were developed and carefully verified by the authors.

## References

[1] W. X. Zhao, K. Zhou, J. Li, T. Tang, X. Wang, Y. Hou, et al., "A Survey of Large Language Models," Frontiers of Computer Science, vol,20, no.12, 2026, pp: 1-40.
[2] H. Touvron, L. Martin, K. Stone, P. Albert, A. Almahairi, Y. Babaei, et al., "Llama 2: Open Foundation and Fine-Tuned Chat Models," arXiv:2307.09288, 2023, pp: 1-77.
[3] OpenAI, "GPT-4 Technical Report," arXiv:2303.08774, 2023, pp: 1-100.
[4] L. Wang, C. Ma, X. Feng, Z. Zhang, et al., "A Survey on Large Language Model Based Autonomous Agents," Frontiers of Computer Science, vol.18, no.6, 2024, pp: 1-26.
[5] S. Yang, C. Fu, S. Zhao, K. Li, X. Sun, T. Xu, E. Chen, "A Survey on Multimodal Large Language Models," National Science Review, vol.11, no.12, 2024, pp: 1-21.
[6] Z. Zhou, X. Chen, E. Li, L. Zeng, K. Luo, and J. Zhang, "Edge Intelligence: Paving the Last Mile of Artificial Intelligence With Edge Computing," Proceedings of the IEEE, vol. 107, no. 8, pp. 1738-1762, 2019.
[7] M. Zhang, X. Shen, J. Cao, Z. Cui, S. Jiang, "EdgeShard: Efficient LLM Inference via Collaborative Edge Computing," IEEE Internet of Things Journal, vol.12, no.10, 2024, pp: 13119-13131.
[8] Y. Sheng, L. Zheng, B. Yuan, Z. Li, M. Ryabinin, et al., "FlexGen: High-Throughput Generative Inference of Large Language Models with a Single GPU," ICML, 2023, Hawaii, USA, pp: 31094 - 31116.
[9] G. Qu, W. Chen, W. Wei, Z. Lin, X. Chen, K. Huang, "Mobile Edge Intelligence for Large Language Models: A Contemporary Survey," IEEE Communications & Tutorials, vol.27, no.6, 2025, pp: 3820-3860.
[10] Y. Zheng, Y. Chen, B. Qian, X. Shi, Y. Shu, J. Chen, "A Review on edge large language models: Design, Execution, and Applications," ACM Computing Surveys, vol.57, no.8, 2025, pp: 1-35.
[11] O. Friha, M.A. Ferrag, B. Kantarci, B. Cakmak, et al., "LLM-based edge intelligence: A Comprehensive survey on Architectures, Applications, Security and Trustworthiness," IEEE Open Journal of the Communication Society, vol.5, 2024, pp: 5799-5856.
[12] G. Cai, R. Tian, L. Yang, Y. Jia, L. Li, J. Wang, "Efficient Inference for edge large language models: A Survey," Tsinghua Science and Technology, vol.31, no.6, 2026, pp: 1365-1380.
[13] N. Shazeer, A. Mirhoseini, K. Maziarz, A. Davis, Q. Le, G. Hinton, and J. Dean, "Outrageously Large Neural Networks: The Sparsely-Gated Mixture-of-Experts Layer," ICLR, 2017, Toulon, France, pp: 1-19.
[14] W. Fedus, B. Zoph, and N. Shazeer, "Switch Transformers: Scaling to Trillion Parameter Models with Simple and Efficient Sparsity," Journal of Machine Learning Research, vol. 23, no. 1, 2022, pp. 5232-5270.
[15] A. Q. Jiang, A. Sablayrolles, A. Mensch, C. Bamford, D. S. Chaplot, D. de las Casas, et al., "Mixtral of Experts," arXiv:2401.04088, 2024, pp: 1-13.
[16] L. Xue, Y. Fu, Z. Lu, L. Mai, M. Marina, "MoE-Infinity: Efficient MoE Inference on Personal Machines with Sparsity-Aware Expert Cache," arXiv:2401.14361, 2024, pp: 1-11.
[17] C. Hwang, W. Cui, Y. Xiong, Z. Yang, Z. Liu, H. Hu, et al., "Tutel:

Adaptive Mixture-of-Experts at Scale," MLSys, 2023, Miami, USA, pp: 1-19.
[18] T. Gale, D. Narayanan, C. Young, and M. Zaharia, "MegaBlocks: Efficient Sparse Training with Mixture-of-Experts," MLSys, 2023, Miami, USA, pp: 1-17.
[19] Z. Jiang, H. Lin, Y. Zhong, Q. Huang, Y. Chen, Z. Zhang, et al., "MegaScale: Scaling Large Language Model Training to More Than 10,000 GPUs," USENIX NSDI, 2024, CA, USA, pp: 745-760.
[20] H. Wang, Q. Zhou, Z. Hong, S. Guo, " D2MoE: Dual Routing and Dynamic Scheduling for Efficient On-Device MoE-based LLM Serving," ACM MobiCom, 2025, Hong Kong, China, pp: 574-588.
[21] Z. Du, S. Li, Y. Wu, et al., "SiDA-MoE: Sparsity-Inspired Data-Aware Serving for Efficient and Scalable Large Mixture-of-Experts Models," MLSys, 2024, pp: 1-15.
[22] A. Borzunov, M. Ryabinin, A. Chumachenko, D. Baranchuk, T. Dettmers, Y. Belkada, et al., "Petals: Collaborative Inference and Fine-tuning of Large Models," ACL, 2023, Toronto, Canda, pp: 558-568.
[23] T. Wu, L. Wang, Z. Wen, X. Zhang, X. Chen, J. Duan, X. Zhang, J. Zuo, "Accelerating Edge Inference for Distributed MoE Models with Latency-Optimized Expert Placement," arXiv:2508.12851, 2025, pp: 1-11.
[24] R. Yi, L. Guo, S. Wei, A. Zhou, S. Wang, M. Xu, "EdgeMoE: Empowering Sparse Large Language Models on Mobile Devices," IEEE Transactions on Mobile Computing, vol.24, no.8, 2025, pp: 7059-7073.
[25] Q. Song, S. Jing, S. Zhang, S. Zhang, C. Huang, " Mixture-of-Experts for Distributed Edge Computing with Channel-Aware Gating Function," IEEE ICC, 2025, Montreal, Canda, pp: 1-6.
[26] Y. Kang, J. Hauswald, C. Gao, A. Rovinski, T. Mudge, J. Mars, and L. Tang, "Neurosurgeon: Collaborative Intelligence Between the Cloud and Mobile Edge," ACM SIGARCH Computer Architecture News, vol.45, no.1, 2017, pp: 615-629.
[27] S. Teerapittayanon, B. McDanel, and H. T. Kung, "Distributed Deep Neural Networks over the Cloud, the Edge and End Devices," IEEE ICDCS, GA, USA, 2017, pp: 1-12.
[28] A. E. Eshratifar, M. S. Abrishami, and M. Pedram, "JointDNN: An Efficient Training and Inference Engine for Intelligent Mobile Cloud Computing Services," IEEE Transactions on Mobile Computing, vol. 20, no. 2, 2021, pp. 565-576.
[29] X. Yuan, N. Li, K. Wei, W. Xu, Q. Chen, H. Chen, S. Guo, "Mobility and Cost Aware Inference Accelerating Algorithm for Edge Intelligence," IEEE Transactions on Mobile Computing, vol.24, no.3, 2025, pp: 1530 - 1549.
[30] Q. Wu, Y. Zhang, C. Yang, J. Sun, "Joint Optimization of Model Partitioning and Resource Allocation for Edge Computing with Intermittently Operating Devices," IEEE ICPADS, Ocean Flower Island, China, 2023, pp: 1-8.
[31] S. Laskaridis, S. I. Venieris, H. Kim, and N. D. Lane, "HAPI: Hardware-Aware Progressive Inference," IEEE/ACM ICCAD, CA, USA, 2020, pp: 1-9.
[32] E. Li, L. Zeng, Z. Zhou, and X. Chen, "Edge AI: On-Demand Accelerating Deep Neural Network Inference via Edge Computing," IEEE Transactions on Wireless Communications, vol. 19, no. 1, 2020, pp. 447-457.
[33] J. Dean, G. Corrado, R. Monga, et al., "Large Scale Distributed Deep Networks," NeurIPS, Lake Tahoe, 2012, pp: 1223-1231.
[34] T. Dettmers, A. Pagnoni, A. Holtzman, and L. Zettlemoyer, "QLoRA: Efficient Finetuning of Quantized LLMs," NeurIPS, New Orleans, USA, 2023, pp: 1-28.
[35] E. Frantar, S. Ashkboos, T. Hoefler, and D. Alistarh, "OPTQ: Accurate Quantization for Generative Pre-trained Transformers," ICLR, Kigali Rwanda, 2023, pp: 1-16.
[36] G. Xiao, J. Lin, M. Seznec, H. Wu, J. Demouth, and S. Han, "SmoothQuant: Accurate and Efficient Post-Training Quantization for Large Language Models," ICML, Hawaii, USA, 2023, pp: 1-13.
[37] J. Lin, J. Tang, H. Tang, S. Yang, X. Dang, and S. Han, "AWQ: Activation-Aware Weight Quantization for LLM Compression and Acceleration," MLSys, Santa Clara, USA, 2024, pp: 1-15.
[38] X. Yuan, N. Li, Q. Chen, W. Xu, A. V. Vasilakos, S. Guo, H. Zhang, "OrderMoE: An expert similarity driven distributed edge MoE inference," arXiv: 2607.17154, 2026, pp: 1-17.